\documentclass[journal]{IEEEtran}

\usepackage{amsfonts}
\usepackage{amssymb}
\usepackage{mathrsfs}
\usepackage{amsmath}
\usepackage{cite}
\usepackage{mathrsfs}
\usepackage{algorithmic}
\usepackage{algorithm}
\usepackage{pgf}
\usepackage{tikz}
\usetikzlibrary{arrows,automata}
\usepackage[latin1]{inputenc}
\usepackage{verbatim}
\makeatletter

\newcommand{\Rmnum}[1]{\expandafter\@slowromancap\romannumeral #1@}
\renewcommand{\algorithmicrequire}{\textbf{Input:}}
\renewcommand{\algorithmicensure}{\textbf{Output:}}
\newcommand{\algorithmicInitialization}{\textbf{Initialization:}}

\makeatother

\newtheorem{thm}{Theorem}
\newtheorem{lemma}[thm]{Lemma}
\newtheorem{eg}{Example}
\newtheorem{prop}{Proposition}
\newtheorem{cor}[thm]{Corollary}
\newtheorem{defn}{Definition}
\newtheorem{remark}{Remark}

\newtheorem{rem-eg}[thm]{Remark and Example}

\newcommand{\p}{{\rho}}
\newcommand{\dt}{{\delta_t}}
\newcommand{\bt}{{\beta_t}}
\newcommand{\f}{\tilde{f}}
\newcommand{\g}{\tilde{g}}
\newcommand{\mL}{\mathcal{L}}
\newcommand{\mF}{\mathcal{F}}
\newcommand{\mP}{\mathcal{P}}
\newcommand{\mO}{\mathcal{O}}
\newcommand{\Rank}{{\mathrm{Rank}}}

\begin{document}
%
\title{Construction of Network Error Correction Codes in Packet Networks
\thanks{This research is supported in part by the National Natural Science Foundation of
China under the Grants 60872025 and 10990011. }}

\author{Xuan~Guang,
        Fang-Wei~Fu,
        and~Zhen Zhang,~\IEEEmembership{Fellow,~IEEE}
\thanks{X. Guang is with the Chern Institute of
Mathematics, Nankai University, Tianjin 300071, P.R. China. Email:
xuanguang@mail.nankai.edu.cn.}
\thanks{F.-W. Fu is with the Chern Institute of
Mathematics and LPMC, Nankai University, Tianjin 300071, P.R. China. Email:
fwfu@nankai.edu.cn.}
\thanks{Z. Zhang is with the Communication Sciences Institute, Ming Hsieh Department of
Electrical Engineering, University of Southern California, Los Angeles,
CA 90089-2565 USA. Email: zhzhang@usc.edu.}}

\markboth{Construction of Network Error Correction Codes in Packet Networks}%
{Guang \MakeLowercase{\textit{et al.}}: Construction of Network Error Correction Codes in Packet Networks}

\maketitle

\begin{abstract}
Recently, network error correction coding (NEC) has been studied extensively. Several bounds in classical coding theory have been extended to network error correction coding, especially the Singleton bound. In this paper, following the research line using the extended global encoding kernels proposed in \cite{zhang-correction}, the refined Singleton bound of NEC can be proved more explicitly. Moreover, we give a constructive proof of the attainability of this bound and indicate that the required field size for the existence of network maximum distance separable (MDS) codes can become smaller further. By this proof, an algorithm is proposed to construct general linear network error correction codes including the linear network error correction MDS codes. Finally, we study the error correction capability of random linear network error correction coding. Motivated partly by the performance analysis of random linear network coding \cite{Ho-etc-random}, we evaluate the different failure probabilities defined in this paper in order to analyze the performance of random linear network error correction coding. Several upper bounds on these probabilities are obtained and they show that these probabilities will approach to zero as the size of the base field goes to infinity. Using these upper bounds, we slightly improve on the probability mass function of the minimum distance of random linear network error correction codes in \cite{zhang-random}, as well as the upper bound on the field size required for the existence of linear network error correction codes with degradation at most $d$.
\end{abstract}

\begin{IEEEkeywords}
Network coding, network error correction coding, the refined Singleton bound, maximum distance separable (MDS) code, random linear network error correction coding, the extended global encoding kernels, network error correction code construction.
\end{IEEEkeywords}

%
\IEEEpeerreviewmaketitle

\section{Introduction}

\IEEEPARstart{N}{etwork} coding was first introduced by Yeung and Zhang in \cite{Zhang-Yeung-1999} and then was profoundly developed by Ahlswede \textit{et al.} \cite{Ahlswede-Cai-Li-Yeung-2000}. In the latter paper \cite{Ahlswede-Cai-Li-Yeung-2000}, the authors showed that by network
coding in network communication, the source node can multicast the information to all sink
nodes at the theoretically maximum rate as the alphabet size
approaches infinity, where the theoretically maximum rate is the smallest minimum cut capacity between the source node and any sink node. Li \textit{et al.} \cite{Li-Yeung-Cai-2003}
indicated that linear network coding with finite alphabet size is
sufficient for multicast. In \cite{Koetter-Medard-algebraic}, Koetter and M$\acute{\textup{e}}$dard presented an algebraic characterization for network coding. Although network coding can achieve the higher information rate than classical routing, Jaggi \textit{et al.} \cite{co-construction} still proposed a deterministic polynomial-time algorithm for constructing a linear network code. Random linear network coding was originally introduced by Ho \textit{et al.} \cite{Ho-etc-random}, and the authors analyzed the performance of random linear network coding by studying the failure probabilities of the codes. Balli, Yan, and
Zhang \cite{zhang-random} improved on the upper bounds on these failure probabilities and then studied the asymptotic  behavior of the failure probability as the field size goes to infinity. Following \cite{zhang-random}, Guang and Fu \cite{Guang-Fu-random} gave some tight or asymptotically tight bounds for two kinds of failure probabilities and also gave the specific network structures in the worst cases.

Network coding has been extensively studied for several years under the assumption that channels of networks are error-free. Unfortunately, all kinds of errors may occur in network communication such as random errors, erasure errors (packet losses), errors in headers and so on. In order to deal with such problems, network error correction coding (NEC) was studied recently. The original idea of network error correction coding was proposed by Yeung and Cai in their conference paper \cite{Yeung-Cai-coorrect} and developed in their recent journal papers \cite{Yueng-Cai-correct-1}\cite{Yueng-Cai-correct-2}. In the latter two papers, the concept of network error correction codes was introduced as a generalization of the classical error correction codes. They also extended some important bounds from classical error correction codes to network error correction codes, such as the Singleton bound, the Hamming bound, and the Gilbert-Varshamov bound. Although the Singleton bound has been given in Cai and Yeung \cite{Yueng-Cai-correct-1}, Zhang\cite{zhang-correction} and Yang \textit{et al.} \cite{Yang-refined-Singleton}\cite{Yang-thesis} presented the refined Singleton bound independently by using the different methods. Yang \textit{et al.} \cite{Yang-weight}\cite{Yang-thesis} developed a framework for characterizing error correction/detection capabilities of network error correction codes. They defined different minimum distances to measure error correction and error detection capabilities, respectively. It followed an interesting discovery that, for nonlinear network error correction codes, the number of the correctable errors can be more than half of the number of the detectable errors. In \cite{zhang-correction}, Zhang defined the minimum distance of linear network error correction codes and introduced the concept of extended global encoding kernels. Using this concept, Zhang proposed linear network error correction codes in packet networks. Besides coherent networks, this scheme is also suitable to non-coherent networks by recording the extended global encoding kernels in the headers of the packets. Moreover, the extended global encoding kernels are used to form the decoding matrices at sink nodes. As well as in \cite{zhang-beyond}, the decoding principles and decoding beyond the error correction capability were studied. The authors further presented several decoding algorithms and analyzed their performance. In addition, Balli, Yan, and Zhang \cite{zhang-random} studied the error correction capability of random linear network error correction codes. They gave the probability mass function of the minimum distance of random linear network error correction codes. For the existence of a network error correction code with degradation, the upper bound on the required field size was proposed.

In \cite{Koetter-correction}, Koetter and Kschischang formulated a different framework for network error correction coding. In their approach, the source message is represented by a subspace of a fixed vector space and a basis of the subspace is injected into the network. This type of network error correction codes is called subspace codes.

In this paper, we follow the research line using the extended global encoding kernels introduced by Zhang in \cite{zhang-packet-meetting}\cite{zhang-correction}. We reprove the refined Singleton bound of the network error correction codes more explicitly by using the concept of the extended global encoding kernels. Similar to the Singleton bound in classical coding theory, the refined Singleton bound is also tight and those linear network error correction codes achieving this bound with equality are called linear network error correction maximum distance separable (MDS) codes, or network MDS codes for short. For network MDS codes, Zhang \cite{zhang-correction} gave an existence proof by an algebraic method. In this paper, we present a constructive proof of the attainability of the refined Singleton bound, and indicate that the required field size for the existence of network MDS codes can become smaller (in some cases much smaller) than the known results. Moreover, by this proof, we design an algorithm for constructing general linear network error correction codes, in particular, network MDS codes.

Matsumoto \cite{Matsumoto-Singleton} and Yang \textit{et al.} \cite{Yang-refined-Singleton} also proposed the algorithms for constructing network MDS codes. The algorithm of Yang \textit{et al.} designs the codebook and the local encoding kernels separately. On the contrary, Matsumoto's algorithm and our algorithm design them together. As noted above, the required field size of our algorithm is smaller. Moreover, compared with Matsumoto's algorithm, our algorithm needs less storages at each sink node. For the decoding, as mentioned in \cite{Matsumoto-Singleton}, the decoding of Matsumoto's algorithm requires exhaustive search by each sink node for all possible information from the source and all possible errors, and our algorithm can make use of the better and faster decoding algorithms proposed by Zhang, Yan, and Balli in a series of papers \cite{zhang-packet-meetting},\cite{zhang-correction}, and \cite{zhang-beyond} such as the brute force decoding algorithm and the fast decoding algorithm. For the case of decoding network error correction codes beyond the error correction capability in packet networks \cite{zhang-beyond}, our algorithm has more advantages because of the use of extended global encoding kernels. We further study the error correction capability of random linear network error correction coding, and analyze the failure probabilities of constructing network MDS codes and general network error correction codes by using random method, as well as the probability mass function of the minimum distance and the required field size.

This paper is divided into 6 sections. In the next section, we introduce the basic notation and definitions in linear network coding and linear network error correction coding, and give some propositions needed in this paper. In Section \Rmnum{3}, we reprove the refine Singleton bound by using the concept of the extended global encoding kernels, and propose a constructive proof to show the attainability of the refined Singleton bound of NEC. Consequently, we also indicate that the required field size for the existence of network MDS codes can become smaller than the known results. Section \Rmnum{4} is devoted to the algorithm for constructing general linear network error correction codes, including network MDS codes. In Section \Rmnum{5}, we analyze the performance of random linear network error correction codes
. The last section summarizes the works done in this paper.

\section{Basic Notation and Definitions}
In this paper, we follow \cite{zhang-correction} in its notation and terminology.
A communication network is defined as a finite acyclic directed
graph $G=(V,E)$, where the vertex set $V$ stands for the set of
nodes and the edge set $E$ represents the set of communication
channels of the network. The node set $V$ consists of three
disjoint subsets $S$, $T$, and $J$, where $S$ is the set of source
nodes, $T$ is the set of sink nodes, and $J=V-S-T$ is the set of internal nodes. Furthermore, a direct edge $e=(i,j)\in E$ represents a channel leading from node $i$ to node $j$. Node $i$
is called the tail of $e$ and node $j$ is called the
head of $e$, written as $i=tail(e)$,
$j=head(e)$, respectively. Correspondingly, the channel $e$ is
called an outgoing channel of $i$ and an incoming channel of $j$. For a
node $i$, define $Out(i)=\{e\in E:e \mbox{ is an outgoing
channel of }i\},\ In(i)=\{e\in E:e \mbox{ is an incoming
channel of }i\}$. In a communication network, if a sequence of channels $(e_1,e_2,\cdots,e_m)$ satisfies $tail(e_1)=i,\ head(e_m)=j$, and $tail(e_{k+1})=head(e_k)$ for $k=1,2,\cdots,m-1$, then we call the sequence $(e_1,e_2,\cdots,e_m)$ a path from node $i$ to node $j$, or equivalently, a path from channel $e_1$ to node $j$. For each channel $e\in E$, there exists a
positive number $R_e$ called the capacity of $e$. We allow the multiple channels between two nodes and
assume reasonably that the capacity of any channel is 1 per unit time. This means that one field symbol can be transmitted over a channel in one unit time.
A cut between node $i$ and node $j$ is a set of channels whose removal disconnects $i$ from $j$. For unit capacity channels, the capacity of a cut can be regarded as the number of channels in the cut, and the minimum of all capacities of cuts between $i$ and $j$ is called the minimum cut capacity between node $i$ and node $j$. A cut between node $i$ and node $j$ is called a minimum cut if its capacity achieves the minimum cut capacity between $i$ and $j$.
Note that there may exist several minimum cuts between $i$ and $j$, but the minimum cut capacity between them is determined.
The source nodes generate messages and transmit them to all sink nodes over the network by network coding.
In the present paper, we consider single source networks, i.e., $|S|=1$,
and the unique source node is denoted by $s$. The source node $s$ has no incoming channels and any sink node has no outgoing
channels, but we use the concept of imaginary incoming channels of the source node $s$ and assume that these imaginary incoming channels provide the source messages to $s$. Let the information rate be $w$ symbols per unit time. Then the source node has $w$ imaginary incoming channels $d_1',d_2',\cdots,d_w'$ and let $In(s)=\{d_1',d_2',\cdots,d_w'\}$. The source messages are $w$ symbols $\underline{\bf{X}}=(X_1,X_2,\cdots,X_w)$ arranged in a row vector where each $X_i$ is an element of base field $\mathcal{F}$.
Assume that they are transmitted to the source node $s$ through the $w$ imaginary
channels in $In(s)$. By using network coding, source messages are
multicast to and decoded at each sink node.

At each node $i\in V-T$, there is an $|In(i)|\times|Out(i)|$ matrix $K_i=(k_{d,e})_{d\in In(i),e\in Out(i)}$ called the local encoding kernel at $i$, where $k_{d,e}\in \mF$ is called the local encoding coefficient for the adjacent pair of channels $(d,e)$. Denote by $U_e$ the message transmitted over the channel $e$. At the source node $s$, assume that the message transmitted over the $i$th imaginary channel is the $i$th source message, i.e., $U_{d_i'}=X_i$. In general, the message $U_e$ is calculated by the formula $U_e=\sum_{d\in In(tail(e))}k_{d,e}U_d$. As we know from \cite{Zhang-book} \cite{Yeung-book}, the global encoding kernel of a channel $e$ is a $w$-dimensional column vector $f_e$ over the base field $\mF$ satisfying $U_e=\underline{\bf{X}}\cdot f_e$. The global encoding kernels can be determined by the local encoding kernels.

In the case that there is an error in a channel $e$, the output of the channel is $\tilde{U}_e=U_e+Z_e$, where $U_e$ is the message that should be transmitted over the channel $e$ and $Z_e\in \mF$ is the error occurred in $e$. We treat $Z_e$ as a message called \textit{error message}. To explain the approach, the extended network was introduced in \cite{zhang-correction} as follows. In the original network $G=(V,E)$, for each channel $e\in E$, an imaginary channel $e'$ is introduced, which is connected to the tail of $e$ to provide error message. This network $\tilde{G}=(\tilde{V},\tilde{E})$ with imaginary channels is called the extended network, where $\tilde{V}=V$ and $\tilde{E}=E\cup E'\cup \{d_1',d_2',\cdots, d_w'\}$ with $E'=\{e': e\in E\}$. Obviously, $|E'|=|E|$. Then a linear network code for the original network can be extended to a linear network code for the extended network by letting $k_{e',e}=1$ and $k_{e',d}=0$ for all $d\in E\backslash\{e\}$. For each internal node $i$ in the extended network, note that $In(i)$ only includes the real incoming channels of $i$, that is, the imaginary channels $e'$ corresponding to $e\in Out(i)$ are not in $In(i)$. But for the source node $s$, we still define $In(s)=\{d_1',d_2',\cdots,d_w'\}$. In order to distinguish two different types of imaginary channels, we call $d_i'\ (1\leq i\leq w)$ the imaginary message channels and $e'$ for $e\in E$ the imaginary error channels. We can also define global encoding kernel $\f_e$ for each $e\in \tilde{E}$ in the extended network. It is a $(w+|E|)$-dimensional column vector and the entries can be indexed by the elements of $In(s)\cup E$. For imaginary message channels $d_i'\ (1\leq i \leq w)$ and imaginary error channels $e'\in E'$, let $\f_{d_i'}=1_{d_i'}$, $\f_{e'}=1_e$, where $1_d$ is a $(w+|E|)$-dimensional column vector which is the indicator function of $d\in In(s)\cup E$. For other global encoding kernels $\f_e, e\in E$, we have recursive formulae:
$$\f_e=\sum_{d\in In(tail(e))}k_{d,e}\f_d+1_e.$$
We call $\f_e$ the extended global encoding kernel of the channel $e\ (e\in E)$ for the original network.
Furthermore, similar to the Koetter-M\'{e}dard Formula\cite{Koetter-Medard-algebraic}, there also exists a formula \cite{zhang-correction}:
$$(\f_e:\ e\in E)=\left( \begin{array}{cc}B\\I \end{array} \right)(I-F)^{-1},$$
where $B=(k_{d,e})_{d\in In(s),e\in E}$ is a $w\times|E|$ matrix with $k_{d,e}=0$ for $e\notin Out(s)$ and $k_{d,e}$ being the local encoding coefficient for $e\in Out(s)$, the system transfer matrix $F=(k_{d,e})_{d\in E, e\in E}$ is an $|E|\times|E|$ matrix with $k_{d,e}$ being the local encoding coefficient for $head(d)=tail(e)$ and $k_{d,e}=0$ for $head(d)\neq tail(e)$, and $I$ is an $|E|\times|E|$ identity matrix.

Let $\underline{\bf{Z}}=(Z_e:\ e\in E)$ be an $|E|$-dimensional row vector with $Z_e\in \mF$ for all $e\in E$. Then $\underline{\bf{Z}}$ is called the error message vector. An error pattern $\p$ is regarded as a set of channels in which errors occur. We call that an error message vector $\underline{\bf{Z}}$ matches an error pattern $\p$, if $Z_e=0$ for all $e\in E\backslash \p$.

For a channel $e\in E$, if there is no error in it, then
$$\tilde{U}_e=(\underline{\bf{X}},\underline{\bf{Z}})\cdot \f_e=(\underline{\bf{X}} ,\underline{\bf{Z}})\cdot (\f_e-1_e)=U_e.$$
If there is an error $Z_e\neq 0$ in channel $e$, then
\begin{align*}
\tilde{U}_e&=U_e+Z_e=(\underline{\bf{X}} ,\underline{\bf{Z}})\cdot (\f_e-1_e)+Z_e\\
&=(\underline{\bf{X}} ,\underline{\bf{Z}})\cdot (\f_e-1_e)+(\underline{\bf{X}} ,\underline{\bf{Z}})\cdot1_e=(\underline{\bf{X}} ,\underline{\bf{Z}})\cdot\f_e.
\end{align*}
At a sink node $t$, the messages $\{\tilde{U}_e:\ e\in In(t)\}$ and the extended global encoding kernels $\{\f_e:\ e\in In(t)\}$ are available. For all messages including information messages and error messages, if they are considered as column vectors, then the above discussions describe linear network error correction coding in packet networks.

First, we need some notation and definitions which either are quoted directly or are extended from Zhang\cite{zhang-correction}.
\begin{defn}[{\cite[Definition 1]{zhang-correction}}]
The matrix
$$\tilde{F}_t=(\f_e:\ e\in In(t))$$
is called the decoding matrix at a sink node $t\in T$. Let
$$\tilde{A}_t=(\tilde{U}_e:\ e\in In(t)).$$
The equation
$$(\underline{\bf{X}} ,\underline{\bf{Z}})\tilde{F}_t=\tilde{A}_t$$
is called the decoding equation at a sink node $t$.
\end{defn}

\begin{defn}
For an error pattern $\p$ and extended global encoding kernels $\f_e,\ e\in E$,
\begin{itemize}
  \item $\f_e^{\p}$ is a $(w+|\p|)$-dimensional column vector obtained from $\f_e=(\f_e(d): d\in In(s)\cup E)$ by removing all entries $\f_e(d),\ d\notin In(s)\cup \p$, and $\f_e^{\p}$ is called the extended global encoding kernel of channel $e$ restricted to the error pattern $\p$.
  \item $f_e^{\p}$ is a $(w+|E|)$-dimensional column vector obtained from $\f_e=(\f_e(d): d\in In(s)\cup E)$ by replacing all entries $\f_e(d),\ d\notin In(s)\cup \p$ by $0$, and $f_e^{\p}$ is also called the extended global encoding kernel of channel $e$ restricted to the error pattern $\p$.
  \item $f_e^{\p^c}$ is a $(w+|E|)$-dimensional column vector obtained from $\f_e=(\f_e(d): d\in In(s)\cup E)$ by replacing all entries $\f_e(d),\ d\in In(s)\cup \p$ by $0$.
\end{itemize}
\end{defn}
Note that $f_e^{\p}+f_e^{\p^c}=\f_e$.

\begin{defn}[{\cite[Defintion 3]{zhang-correction}}]
Define
$$\Delta(t,\p)=\{(\underline{\bf 0},\underline{\bf{Z}})\tilde{F}_t:\ \mbox{all }\underline{\bf{Z}} \mbox{ matching the error pattern }\p \}$$
where $\underline{\bf0}$ is a $w$-dimensional zero row vector, and $\underline{\bf{Z}}$ is an $|E|$-dimensional row vector matching the error pattern $\p$; and
$$\Phi(t)=\{(\underline{\bf{X}}, \underline{\bf0})\tilde{F}_t:\ \underline{\bf{X}}\in \mF^w \}.$$
We call $\Delta(t,\p)$ the error space of error pattern $\p$ and $\Phi(t)$ the message space.
\end{defn}

Let $L$ be a collection of vectors in a linear space. $\langle L \rangle$ represents the subspace spanned by the vectors in $L$. In fact, if we use $row_t(d),\ d\in In(s)\cup E$ to denote the row vectors of the decoding matrix $\tilde{F}_t$, then $\Delta(t,\p)=\langle \{ row_t(d):\ d\in \p \} \rangle$ and $\Phi(t)=\langle \{ row_t(d):\ d\in In(s) \} \rangle$.

\begin{defn}[{\cite[Definition 4]{zhang-correction}}]
We say that an error pattern $\p_1$ is dominated by another error pattern $\p_2$ with respect to a sink node $t$ if $\Delta(t,\p_1)\subseteq \Delta(t,\p_2)$ for any linear network code. This relation is denoted by $\p_1\prec_t\p_2$.
\end{defn}

\begin{defn}[{\cite[Definition 5]{zhang-correction}}]
The rank of an error pattern $\p$ with respect to a sink node $t$ is defined by
$$rank_t(\p)=\min\{|\p'|:\ \p\prec_t\p'\}.$$
\end{defn}

In order to understand the concept of rank of an error pattern better, we give the following proposition. This proposition is a slight and necessary modification of {\cite[Lemma 1]{zhang-correction}}.
\begin{prop}\label{prop_error_pattern}
For an error pattern $\p$, introduce a source node $s_{\p}$. Let $\p=\{ e_1,e_2,\cdots,e_l \}$ where $e_j\in In(i_j)$ for $1\leq j \leq l$ and define $e_j'=(s_{\p},i_j)$. Replace each $e_j$ by $e_j'$ on the network, that is, add $e_1',e_2',\cdots,e_l'$ on the network and delete $e_1,e_2,\cdots,e_l$ from the network. Then the rank of the error pattern $\p$ with respect to a sink node $t$ is equal to the minimum cut capacity between $s_{\p}$ and $t$.
\end{prop}
\begin{IEEEproof}
It is similar to the proof in \cite{zhang-correction}, and, therefore, omitted.
\end{IEEEproof}
\begin{defn}[{\cite[Definition 6]{zhang-correction}}]
A linear network error correction code is called a regular code if for any $t\in T$, $\dim(\Phi(t))=w$.
\end{defn}

\begin{defn}[{\cite[Definition 7]{zhang-correction}}]
The minimum distance of a regular network error correction code at a sink node $t$ is defined by
$$d_{\min}^{(t)}=\min\{ rank_t(\p):\ \dim(\Delta(t,\p)\cap \Phi(t))>0 \}.$$
\end{defn}

For the minimum distance above, we give the following proposition.

\begin{prop}\label{prop_distance}
For the minimum distance of a regular network error correction code at a sink node $t$, there exist the following equalities:
\begin{align}
d_{\min}^{(t)}&=\min\{ rank_t(\p):\ \Delta(t,\p)\cap \Phi(t)\neq\{\underline{0}\} \}\label{eq_1}\\
              &=\min\{ |\p|:\ \Delta(t,\p)\cap \Phi(t)\neq \{\underline{0}\} \}\label{eq_2}\\
              &=\min\{ \dim(\Delta(t,\p)):\ \Delta(t,\p)\cap \Phi(t)\neq \{\underline{0}\} \}.\label{eq_3}
\end{align}
\end{prop}
\begin{IEEEproof}
We define the set of error patterns $\Pi=\{\p: \Delta(t,\p)\cap \Phi(t)\neq\{\underline{0}\}\}$. Then one has
$$(\ref{eq_1})=\min_{\p\in \Pi}rank_t(\p),\ (\ref{eq_2})=\min_{\p\in \Pi}|\p|,\ (\ref{eq_3})=\min_{\p\in \Pi}\dim(\Delta(t,\p)).$$
Since $\dim(\Delta(t,\p))\leq rank_t(\p) \leq |\p|$ for any error pattern $\p\subseteq E$, it follows that
$$\min_{\p\in \Pi}\dim(\Delta(t,\p))\leq \min_{\p\in \Pi}rank_t(\p)\leq \min_{\p\in \Pi}|\p|.$$

In view of the inequalities above, it is enough to prove $\min_{\p\in \Pi}|\p|\leq \min_{\p\in \Pi}\dim(\Delta(t,\p))$.
Let $\p'\in \Pi$ be an error pattern satisfying
$$\dim(\Delta(t,\p'))=\min_{\p\in \Pi}\dim(\Delta(t,\p)).$$
Assume that $\p'=\{e_1,e_2,\cdots,e_l\}$, which means $\Delta(t,\p')=\langle \{ row_t(e_i): 1\leq i \leq l \} \rangle$.
For $\{ row_t(e_i): 1\leq i \leq l \}$, let its maximum independent vector set be
$\{ row_t(e_{i_j}): 1\leq j \leq m \}$, where $m=\dim(\Delta(t,\p'))\leq l$. Set
$\p_1=\{e_{i_j}: 1\leq j \leq m\}$. This implies that
$$|\p_1|=\dim(\Delta(t,\p_1))=\dim(\Delta(t,\p'))$$ and
$$\Delta(t,\p_1)\cap \Phi(t)=\Delta(t,\p')\cap \Phi(t)\neq \{\underline{0}\}.$$
Therefore,
$$\min_{\p\in \Pi}|\p|\leq |\p_1|=\dim(\Delta(t,\p'))=\min_{\p\in \Pi}\dim(\Delta(t,\p)).$$
The proof is completed.
\end{IEEEproof}

In this paper, we always use $w$ to denote the information rate and $C_t$ to denote the minimum cut capacity between the unique source node $s$ and sink node $t$, and define $\dt=C_t-w$ which is called the redundancy of sink node $t$.

\section{The Refined Singleton Bound of NEC and The Network MDS Codes}
By using the concept of the extended global encoding kernels, we can reprove the refined Singleton bound of NEC. First, we give the following lemma.

\begin{lemma}\label{lem_Singleton}
For a regular linear network error correction code, let a channel set $\{e_1,e_2,\cdots,e_{C_t}\}$ be a minimum cut between $s$ and $t$ with an upstream-to-downstream order $e_1\prec e_2\prec \cdots\prec e_{C_t}$ and let an error pattern $\p=\{e_{w}, e_{w+1}, \cdots, e_{C_t}\}$. Then $\Phi(t)\cap \Delta(t,\p)\neq \{\underline{0}\}$.
\end{lemma}
\begin{IEEEproof}
Let $\underline{\bf{X}}$ and $\underline{\bf{Z}}$ represent the source message vector and the error message vector, respectively. Then, for each channel $e\in E$, we have $\tilde{U}_e=(\underline{\bf{X}},\underline{\bf{Z}})\cdot\f_e$, where $\tilde{U}_e$ is the output of $e$.
Let $\tilde{U}_{e_1}=\tilde{U}_{e_2}=\cdots=\tilde{U}_{e_{w-1}}=0$.
Since $\Rank((\f_{e_1}\ \f_{e_2}\ \cdots\ \f_{e_{w-1}}))$ is at most $(w-1)$, there exists a nonzero message vector $\underline{\bf{X}}_1$ and an error message vector $\underline{\bf{Z}}_1=\underline{\bf{0}}$ such that
\begin{align*}
&(\underline{\bf{X}}_1,\underline{\bf{Z}}_1)\cdot(\f_{e_1}\ \f_{e_2}\ \cdots\ \f_{e_{w-1}})\\
=&(\underline{\bf{X}}_1,\underline{\bf{0}})\cdot(\f_{e_1}\ \f_{e_2}\ \cdots\ \f_{e_{w-1}})\\
=&(\tilde{U}_{e_1}\ \tilde{U}_{e_2}\ \cdots\ \tilde{U}_{e_{w-1}})=\underline{\bf{0}} .
\end{align*}
Moreover, as this code is regular, this implies
$$(\underline{\bf{X}}_1,\underline{\bf{0}})\cdot(\f_{e_1}\ \f_{e_2}\ \cdots\ \f_{e_{C_t}})=
(\tilde{U}_{e_1}\ \tilde{U}_{e_2}\ \cdots\ \tilde{U}_{e_{C_t}})\neq \underline{\bf{0}}.$$
Assume the contrary, i.e., $(\tilde{U}_{e_1}\ \tilde{U}_{e_2}\ \cdots\ \tilde{U}_{e_{C_t}})=\underline{\bf{0}}$. And note that $\{e_1,e_2,\cdots,e_{C_t}\}$ is a minimum cut between $s$ and $t$ and $\underline{\bf{Z}}_1=\underline{\bf{0}}$. It follows that
$$\tilde{A}_t=(\tilde{U}_e: e\in In(t))=\underline{\bf{0}},$$
which implies that $(\underline{\bf{X}}_1,\underline{\bf{0}})\tilde{F}_t=\underline{\bf{0}}$ from the decoding equation $(\underline{\bf{X}}_1,\underline{\bf{Z}}_1)\tilde{F}_t=\tilde{A}_t$. Therefore, the equality $\underline{\bf{X}}_1=\underline{\bf{0}}$ follows from $\dim(\Phi(t))=w$ because the linear network error correction code considered is regular. This contradicts our assumption $\underline{\bf{X}}_1\neq\underline{\bf{0}}$.

On the other hand, there exists another source message vector $\underline{\bf{X}}_2={\bf\underline{0}}$ and another error message vector $\underline{\bf{Z}}_2$ matching the error pattern $\p=\{e_{w},e_{w+1},\cdots,e_{C_t}\}$, such that
$$(\underline{\bf{X}}_2,\underline{\bf{Z}}_2)\cdot(\f_{e_1}\ \f_{e_2}\ \cdots\ \f_{e_{C_t}})=
(\tilde{U}_{e_1}\ \tilde{U}_{e_2}\ \cdots\ \tilde{U}_{e_{C_t}}).$$
And note that $\underline{\bf{Z}}_2\neq \underline{\bf 0}$ because $(\tilde{U}_{e_1}\ \tilde{U}_{e_2}\ \cdots\ \tilde{U}_{e_{C_t}})\neq \underline{\bf{0}}$.
In fact, since $e_{w}\prec e_{w+1}\prec \cdots\prec e_{C_t}$, for $e\in \p$, we can set sequentially:
$$Z_e=\tilde{U}_e-\sum_{d\in In(tail(e))}k_{d,e}\tilde{U}_d',$$
where $\tilde{U}_d'$ is the output of channel $d$ in this case.


Therefore, it follows that
$$(\underline{\bf{X}}_1,\underline{\bf{0}})\cdot \tilde{F}_t =(\underline{\bf{0}},\underline{\bf{Z}}_2)\cdot \tilde{F}_t .$$
And note that $\underline{\bf{Z}}_2$ matches the error pattern $\p$. It is shown that
$\Phi(t)\cap \Delta(t,\p)\neq \{\underline{0}\}$. The lemma is proved.
\end{IEEEproof}

\begin{thm}[The Refined Singleton Bound]\label{thm_ref_singleton_b}
Let $d_{\min}^{(t)}$ be the minimum distance of a regular linear network error correction code at a sink node $t\in T$. Then
$$d_{\min}^{(t)}\leq \delta_t+1.$$
\end{thm}
\begin{remark}
Conventionally, if a regular network error correction code $\mathbf{C}$ satisfies the refined Singleton bound with equality, that is, $d_{\min}^{(t)}=\delta_t+1$ for each $t\in T$, then this code $\mathbf{C}$ is called network error correction maximum distance separable (MDS) code, or network MDS codes for short.
\end{remark}

It is not hard to see that Theorem \ref{thm_ref_singleton_b} is an obvious consequence of Proposition \ref{prop_distance} and Lemma \ref{lem_Singleton}. Now, we give a constructive proof to show that the refined Singleton bound is tight.
First, we need the following lemma from \cite{zhang-correction}.
Define $R_t(\delta_t)$ as the set of the error patterns $\p$ satisfying $|\p|=rank_t(\p)=\delta_t$, that is,
$$R_t(\delta_t)=\{\mbox{error pattern}\ \p:\ |\p|=rank_t(\p)=\delta_t\}.$$

\begin{lemma}\label{lem_path}
For each $t\in T$ and any error pattern $\p\in R_t(\delta_t)$, there exist $(w+\delta_t)$ channel disjoint paths from either $s$ or $\p$ to $t$, and the $(w+\delta_t)$ paths satisfy the properties that
\begin{enumerate}
  \item there are exactly $\delta_t$ paths from $\p$ to $t$, and $w$ paths from $s$ to $t$;
  \item these $\delta_t$ paths from $\p$ to $t$ start with the different channels in $\p$.
\end{enumerate}
\end{lemma}

Furthermore, in Lemma \ref{lem_path}, assign $w$ imaginary message channels $d_1',d_2',\cdots,d_w'$ to the $w$ paths from $s$ to $t$,
and assign $\dt$ imaginary error channels $e', e\in \p$ to the $\dt$ paths from $\p$ to $t$, i.e., for each $e\in \p$, assign $e'$ to the path from $e$ to $t$. This leads to the following corollary.

\begin{cor}\label{coro_path}
For each $t\in T$ and any error pattern $\p\in R_t(\delta_t)$, there exist $(w+\delta_t)$ channel disjoint paths from either $In(s)=\{d_1',d_2',\cdots,d_w'\}$ or $\p'=\{e': e\in \p\}$ to $t$, and the $(w+\dt)$ paths satisfy the properties that
\begin{enumerate}
  \item there are exactly $\delta_t$ paths from $\p'$ to $t$, and $w$ paths from $In(s)$ to $t$;
  \item these $\dt$ paths from $\p'$ to $t$ start with the distinct channels in $\p'$ and for each path, if it starts with $e'\in \p'$, then it passes through $e\in \p$.
\end{enumerate}
\end{cor}

\begin{thm}\label{thm_MDS}
If $|\mathcal{F}|\geq \sum_{t\in T}|R_t(\delta_t)|$, then there exist linear network error correction MDS codes, i.e., for all $t\in T$,
$$d_{\min}^{(t)}=\dt+1.$$
\end{thm}
\begin{IEEEproof}
Let $G=\{V,E\}$ be a single source multicast network, where $s$ is the single source, $T$ is the set of sink nodes, $J=V-\{s\}-T$ is the set of  internal nodes, and $E$ represents the set of channels in $G$.
Let $\tilde{G}=(\tilde{V},\tilde{E})$ be the extended network of $G$. For each $t\in T$ and each $\rho\in R_t(\dt)$, $\mP_{t,\p}$ denotes the set of $(w+\dt)$ channel disjoint paths satisfying Corollary \ref{coro_path}.
Denote by $E_{t,\p}$ the set of all channels on paths in $\mathcal{P}_{t,\p}$.

Now, we define a dynamic set of channels $CUT_{t,\p}$ for each $t\in T$ and each $\p\in R_t(\dt)$, and initialize
$$CUT_{t,\p}=In(s)\cup \p'=\{d_1',d_2',\cdots,d_w'\} \cup \{e':\ e\in \p\},$$
where $e'$ is the imaginary error channel corresponding to $e$. Initialize $\f_d=\underline{0}$ for all $d\in E$ and $\f_d=1_d$ for all $d\in In(s)\cup E'$. Naturally, we are interested in $\{\tilde{f}_{d}: d\in CUT_{t,\p}\}$.

For any subset $B\subseteq In(s)\cup E'\cup E$, define
\begin{align*}
&\tilde{\mL}(B)=\langle \{ \f_e: e\in B \} \rangle, \tilde{\mL}^{\p}(B)=\langle \{ \f^{\p}_e: e\in B \} \rangle,
\mbox{ and}\\
&\mL^{\p}(B)=\langle \{ f^{\p}_e: e\in B \} \rangle, \mL^{\p^c}(B)=\langle \{ f^{\p^c}_e: e\in B \} \rangle.
\end{align*}

For $CUT_{t,\p}$, note that the initial set is $CUT_{t,\p}=In(s)\cup \p'$, which means
\begin{align*}
\tilde{\mL}(CUT_{t,\p})&=\langle \{\f_d: d\in In(s)\cup \p'\} \rangle\\
&=\langle \{1_d: d\in In(s)\cup \{e':e\in\p\} \}\rangle.
\end{align*}
Thus $\begin{pmatrix}\f_d^{\p}: d\in CUT_{t,\p}\end{pmatrix}=\begin{pmatrix}\f_d^{\p}: d\in In(s)\cup\p'\end{pmatrix}$ is an identity matrix of size $(w+\dt)\times(w+\dt)$. That is, \linebreak $\Rank(\begin{pmatrix}\f_d^{\p}: d\in In(s)\cup\p'\end{pmatrix})=w+\dt$ or $\dim(\tilde{\mL}^{\p}(CUT_{t,\p}))\\=w+\dt$.

Next, we will update $CUT_{t,\p}$ in the topological order of all nodes until $CUT_{t,\p}\subseteq In(t)$.

For each $i\in V$, consider all channels $e\in Out(i)$ in arbitrary order.
For each $e\in Out(i)$, if $e\notin \cup_{t\in T}\cup_{\p\in R_t(\dt)}E_{t,\p}$, let $\f_e=1_e$, and all $CUT_{t,\p}$ remain unchanged.
Otherwise $e\in \cup_{t\in T}\cup_{\p\in R_t(\dt)}E_{t,\p}$, i.e., $e\in E_{t,\p}$ for some $t\in T$ and $\p\in R_t(\dt)$. In $\mP_{t,\p}$, we use $e(t,\p)$ to denote the previous channel of $e$ on the path which $e$ locates on.
Choose
\begin{align}\label{1}
&\g_e\in  \tilde{\mL}(In(i)\cup \{e'\})\backslash\\
&\cup_{t\in T}\cup_{\p\in R_t(\dt):\atop e\in E_{t,\p}}[\mL^{\p}(CUT_{t,\p}\backslash \{e(t,\p)\})+\mL^{\p^c}(In(i)\cup \{e'\})],\nonumber
\end{align}
where the addition ``$+$'' represents the sum of two vector spaces. Further, let
$$\f_e=\left\{\begin{array}{ll}
\tilde{g}_e+1_e&\ \mbox{if } \tilde{g}_e(e)=0,\\
\tilde{g}_e(e)^{-1}\cdot\tilde{g}_e&\ \mbox{otherwise}.
\end{array}\right.$$
For those $CUT_{t,\p}$ satisfying $e\in E_{t,\p}$, update $CUT_{t,\p}=\{CUT_{t,\p}\backslash\{e(t,\p)\}\}\cup\{e\}$; and for others, $CUT_{t,\p}$ remain unchanged.

Updating all channels in $E$ by the same method, one can see that all $\f_e, e\in E$ are well-defined and, finally, $CUT_{t,\p}\subseteq In(t)$ for all $t\in T$ and $\p\in R_t(\dt)$.

To complete the proof, we only need to prove the following two conclusions:
\begin{enumerate}
  \item For each $t\in T$, $d_{\min}^{(t)}=\dt+1$.
  \item There exists nonzero column vector $\tilde{g}_e$ satisfying (\ref{1}).
\end{enumerate}

{\bf\textit{The proof of 1):}} We will indicate that all $CUT_{t,\p}$ satisfy $\dim(\tilde{\mL}^{\p}(CUT_{t,\p}))=w+\dt$ during the whole updating process by induction.

Assume that all channels before $e$ have been updated and $\dim(\tilde{\mL}^{\p}(CUT_{t,\p}))=w+\dt$ for each $CUT_{t,\p}$. Now, we take the channel $e$ into account. Since we choose
\begin{align*}
&\g_e\in \tilde{\mL}(In(i)\cup \{e'\})\backslash\\
&\cup_{t\in T}\cup_{\p\in R_t(\dt):\atop e\in E_{t,\p}}[\mL^{\p}(CUT_{t,\p}\backslash \{e(t,\p)\})+\mL^{\p^c}(In(i)\cup \{e'\})],
\end{align*}
it follows that $\g_e^{\p}$ and $\{\f_d^{\p}: d\in CUT_{t,\p}\backslash\{e(t,\p)\}\}$ are linearly independent for any $CUT_{t,\p}$ with $e\in E_{t,\p}$.
Conversely, suppose that $\g_e^{\p}$ and $\{\f_d^{\p}: d\in CUT_{t,\p}\backslash\{e(t,\p)\}\}$ are linearly dependent. This means that $g_e^{\p}$ is a linear combination of vectors in $\{f_d^{\p}: d\in CUT_{t,\p}\backslash\{e(t,\p)\}\}$. And $g_e^{\p^c}$ is a linear combination of vectors in $\{ f_d^{\p^c}: d\in In(i)\cup \{e'\} \}$ because of $\g_e\in \tilde{\mL}(In(i)\cup \{e'\})$. Therefore, $\g_e=g_e^{\p}+g_e^{\p^c}$ is a linear combination of vectors in
$$\{ f_d^{\p}:d\in CUT_{t,\p}\backslash\{e(t,\p)\} \}\cup \{ f_d^{\p^c}: d\in In(i)\cup \{e'\} \}.$$
This is a contradiction to the choice of $\g_e$.

In the following, we will show that $\f_e^{\p}$ and $\{ \f_d^{\p}: d\in CUT_{t,\p}\backslash\{e(t,\p)\} \}$ are also linearly independent.
\begin{itemize}
  \item If $\g_e(e)\neq0$, then, since $\g_e^{\p}$ and $\{ \f_d^{\p}: d\in CUT_{t,\p}\backslash\{e(t,\p)\} \}$ are linearly independent, $\f_e^{\p}=\g_e(e)^{-1}\cdot\g_e^{\p}$ and $\{ \f_d^{\p}: d\in CUT_{t,\p}\backslash\{e(t,\p)\} \}$ are also linearly independent.

  \item Otherwise $\g_e(e)=0$. We claim that $e\notin \p$. Assume the contrary, i.e., $e\in \p$. Thus $e(t,\p)=e'$ which means $\f_{e(t,\p)}=1_e$ and $\f_d(e)=0$ for all $d\in CUT_{t,\p}\backslash\{e(t,\p)\}$. Together with $\g_e(e)=0$ and $\dim(\tilde{\mL}^{\p}(CUT_{t,\p}))=w+\dt$, it follows that $\g_e^{\p}$ is a linear combination of vectors in $\{\f_d^{\p}: d\in CUT_{t,\p}\backslash\{e(t,\p)\} \}$. This implies that $\g_e\in \mL^{\p}(CUT_{t,\p}\backslash\{e(t,\p)\})+\mL^{\p^c}(In(i)\cup\{e'\})$, which leads to a contradiction. Hence, in view of $e \notin \p$, one obtains $\g_e^{\p}=\f_e^{\p}$, which implies that $\f_e^{\p}$ and $\{ \f_d^{\p}: d\in CUT_{t,\p}\backslash\{e(t,\p)\} \}$ are linearly independent.
\end{itemize}

Finally, after all updates, we have $CUT_{t,\p}\subseteq In(t)$ for each $t\in T$ and each $\p\in R_t(\dt)$, and $\Rank(\begin{pmatrix}\f_e^{\p}: e\in CUT_{t,\p}\end{pmatrix})=w+\dt$. As the matrix $\begin{pmatrix}\f_e^{\p}: e\in CUT_{t,\p}\end{pmatrix}$ is a submatrix of $\tilde{F}_t^{\p}\triangleq \begin{pmatrix}\f^{\p}_e: e\in In(t)\end{pmatrix}$ with the same number of rows, it follows that $\Rank(\tilde{F}_t^{\p})=w+\dt$, i.e., $\Phi(t)\cap \Delta(t,\p)=\{\underline{0}\}$.

For each error pattern $\eta\subseteq E$ satisfying $rank_t(\eta)<\dt$, there exists an error pattern $\p\in R_t(\dt)$ such that $\eta \prec_t \p$ from Proposition \ref{prop_error_pattern}. This implies that $\Delta(t,\eta)\subseteq \Delta(t,\p)$,
and thus,
$$\Phi(t)\cap\Delta(t,\eta)\subseteq \Phi(t)\cap\Delta(t,\p)=\{\underline{0}\}.$$
Now, we can say that $d_{\min}^{(t)}\geq \dt+1$ for all $t\in T$, which, together with $d_{\min}^{(t)}\leq \dt+1$ from Theorem \ref{thm_ref_singleton_b}, shows that $d_{\min}^{(t)}=\dt+1$ for all $t\in T$.

{\bf \textit{The proof of 2):}} We just need to prove that if $|\mF|\geq \sum_{t\in T}|R_t(\dt)|$, then
\begin{multline*}
\Big|\tilde{\mL}(In(i)\cup \{e'\})\backslash \hfill \\
\cup_{t\in T}\cup_{\p\in R_t(\dt):\atop e\in E_{t,\p}}[\mL^{\p}(CUT_{t,\p}\backslash \{e(t,\p)\})+\mL^{\p^c}(In(i)\cup \{e'\})]\Big|\\
>0.\hfill
\end{multline*}

Let $\dim(\tilde{\mL}(In(i)\cup\{e'\}))=k$.
For each $t\in T$ and $\p\in R_t(\dt)$, if $e\in E_{t,\p}$, then $e(t,\p)\in In(i)\cup \{e'\}$, i.e., $\f_{e(t,\p)}\in \tilde{\mL}(In(i)\cup \{e'\})$. Moreover, we know $\f_{e(t,\p)}^{\p}\notin \tilde{\mL}^{\p}(CUT_{t,\p}\backslash\{e(t,\p)\})$, i.e., $f_{e(t,\p)}^{\p} \notin \mL^{\p}(CUT_{t,\p}\backslash \{e(t,\p)\})$, and $f_{e(t,\p)}^{\p}\notin \mL^{\p^c}(In(i)\cup \{e'\})$. Together with $f_{e(t,\p)}^{\p^c}\in \mL^{\p^c}(In(i)\cup \{e'\})$ and $\f_{e(t,\p)}=f_{e(t,\p)}^{\p}+f_{e(t,\p)}^{\p^c}$, this implies that $$\f_{e(t,\p)}\notin \mL^{\p}(CUT_{t,\p}\backslash \{e(t,\p)\})+\mL^{\p^c}(In(i)\cup \{e'\}).$$ Therefore,
\begin{multline}\label{dim}
\dim\big(\tilde{\mL}(In(i)\cup \{e'\})\cap\\
     [\mL^{\p}(CUT_{t,\p}\backslash\{e(t,\p)\})+\mL^{\p^c}(In(i)\cup \{e'\})]\big)\leq k-1.
\end{multline}
Consequently,
\begin{align}
&\big|\tilde{\mL}(In(i)\cup\{e'\})\backslash\cup_{t\in T}\cup_{\p\in R_t(\dt):\atop e\in E_{t,\p}}\nonumber\\
&[\mL^{\p}(CUT_{t,\p}\backslash \{e(t,\p)\})+\mL^{\p^c}(In(i)\cup \{e'\})]\big|\nonumber\\
=&\big|\tilde{\mL}(In(i)\cup \{e'\})|-\big|\tilde{\mL}(In(i)\cup \{e'\})\cap\{\cup_{t\in T}\cup_{\p\in R_t(\dt):\atop e\in E_{t,\p}}\nonumber\\
&[\mL^{\p}(CUT_{t,\p}\backslash \{e(t,\p)\})+\mL^{\p^c}(In(i)\cup \{e'\})]\} \big|\label{cap}\\
>&|\mF|^k-\sum_{t\in T}\sum_{\p\in R_t(\dt)}|\mF|^{k-1}\label{sum}\\
\geq&|\mF|^{k-1}[|\mF|-\sum_{t\in T}|R_t(\dt)|\ ]\geq0,\nonumber
\end{align}
where the last step follows from $|\mF|\geq \sum_{t\in T}|R_t(\dt)|$.
For the inequality $(\ref{cap})>(\ref{sum})$, it is readily seen from (\ref{dim}) that $(\ref{cap})\geq(\ref{sum})$. It suffices to show $(\ref{cap})>(\ref{sum})$. It is not difficult to obtain that $(\ref{cap})=(\ref{sum})$, i.e.,
\begin{multline*}
\big|\tilde{\mL}(In(i)\cup \{e'\})\cap\\
\{\cup_{t\in T}\cup_{\p\in R_t(\dt):\atop e\in E_{t,\p}}[\mL^{\p}(CUT_{t,\p}\backslash \{e(t,\p)\})+\mL^{\p^c}(In(i)\cup \{e'\})]\} \big|\\
=\sum_{t\in T}\sum_{\p\in R_t(\dt)}|\mF|^{k-1}\hfill
\end{multline*}
if and only if $|T|=1,\ |R_t(\dt)|=1$ and
\begin{multline*}
\dim(\tilde{\mL}(In(i)\cup\{e'\})\cap\\
     [\mL^{\p}(CUT_{t,\p}\backslash \{e(t,\p)\})+\mL^{\p^c}(In(i)\cup \{e'\})])=k-1
\end{multline*}
with $e\in E_{t,\p}$, where $R_t(\dt)=\{\p\}$.
However, it is impossible that $|R_t(\dt)|=1$ because $\dt<C_t$.
The proof is completed.
\end{IEEEproof}

According to the known results, for the existence of the network error correction MDS codes, the size of the required base field is at least $\sum_{t\in T}{|E|\choose \dt}$. By Theorem \ref{thm_MDS}, we can say that $\sum_{t\in T}|R_t(\dt)|$ is enough.

For any channel $e\in E$, if there exists a path from $e$ to sink node $t$, then we call that $e$ is connective with $t$.
\begin{lemma}\label{lem_field_size}
Let $E_t$ be the set of channels which are connective with sink node $t\in T$. Then
$$\sum_{t\in T}|R_t(\dt)|\leq \sum_{t\in T}{|E_t|\choose \dt} \leq \sum_{t\in T}{|E|\choose \dt}.$$
Moreover, the necessary condition of the second inequality holding with equality is that there exists only one sink node in the network, i.e., $|T|=1$.
\end{lemma}
\begin{IEEEproof}
Both inequalities are clear, and we will only consider the necessary condition of the second inequality holding with equality. Suppose that there are more than one sink node, and let $t$ and $t'$ be two distinct sink nodes. Obviously, there exists a channel $e$ with $head(e)=t'$. That is, $e$ is not connective with sink node $t$. This implies that $|E_t|<|E|$, and thus
${|E_t|\choose \dt}<{|E|\choose \dt}$, which shows that $\sum_{t\in T}{|E_t|\choose \dt}< \sum_{t\in T}{|E|\choose \dt}$. The lemma is proved.
\end{IEEEproof}

From Theorem \ref{thm_MDS} and Lemma \ref{lem_field_size}, we get the following corollary.
\begin{cor}
If $|\mathcal{F}|\geq \sum_{t\in T}{|E_t| \choose \dt}$, then there exist linear network error correction MDS codes, i.e., for all $t\in T$,
$$d_{\min}^{(t)}=\delta_t+1.$$
\end{cor}

\begin{eg}
Let $G$ be a combination network \cite[p.450]{Yeung-book}\cite[p.26]{Zhang-book} with $N=6$ and $k=4$. That is, $G$ is a single source multicast network, where there are $N=6$ internal nodes, and one and only one channel from the source node $s$ to each internal node. Arbitrary $k=4$ internal nodes are connective with one and only one sink node, which implies that there are total ${6\choose 4}=15$ sink nodes. Thus, for $G$, we know that $|J|=6$, $|T|={6 \choose 4}=15$, and $|E|=6+4\times{6\choose 4}=66$. It is evident that the minimum cut capacity $C_t$ between $s$ and any sink node $t$ is $4$. For example, Fig. \ref{fig_cn} shows a combination network with $N=3,k=2$.
\begin{figure}[!htb]
\centering
\begin{tikzpicture}
[->,>=stealth',shorten >=1pt,auto,node distance=2cm,
                    thick]
  \tikzstyle{every state}=[fill=none,draw=black,text=black,minimum size=7mm]
  \tikzstyle{place}=[fill=none,draw=white,minimum size=0.1mm]
  \node[state]         (s)                 {$s$};
  \node[state]         (i_2)[below of=s]   {$i_2$};
  \node[state]         (i_1)[left of=i_2]  {$i_1$};
  \node[state]         (i_3)[right of=i_2] {$i_3$};
  \node[state]         (t_1)[below of=i_1] {$t_1$};
  \node[state]         (t_2)[below of=i_2] {$t_2$};
  \node[state]         (t_3)[below of=i_3] {$t_3$};
\path
(s) edge           node {} (i_1)
    edge           node {} (i_2)
    edge           node {} (i_3)
(i_1) edge node{}(t_1)
      edge node{}(t_2)
(i_2) edge node{}(t_1)
      edge node{}(t_3)
(i_3) edge node{}(t_2)
      edge node{}(t_3);
\end{tikzpicture}
\caption{Combination Network with $N=3,k=2$.}
\label{fig_cn}
\end{figure}
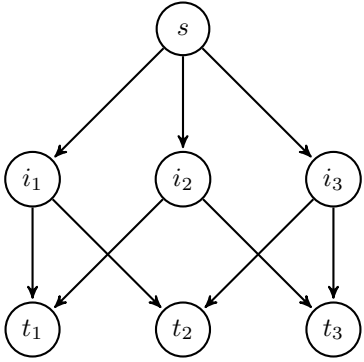
Furthermore, let the information rate be $w=2$, and thus $\dt=2$ for each $t\in T$.
Therefore, $|R_t(\dt)|=|R_t(2)|=4\times{4 \choose 2}=24$ for each $t\in T$, and $\sum_{t\in T}|R_t(\dt)|=15\times24=360$. Nevertheless, $\sum_{t\in T}{|E_t|\choose \dt}=15\times{8\choose 2}=420$ and $\sum_{t\in T}{|E|\choose \dt}=15\times{66\choose 2}=32175$.
\end{eg}

Now, we take into account the general network error correction codes, and give the following theorem.
\begin{thm}\label{thm_general}
For any nonnegative integers $\bt$ with $\bt\leq\dt$ for each $t\in T$, if $|\mathcal{F}|\geq \sum_{t\in T}|R_t(\bt)|$, then there exist linear network error correction codes satisfying for all $t\in T$,
$$d_{\min}^{(t)}\geq\bt+1,$$
where $R_t(\bt)$ is the set of error patterns $\p$ satisfying $|\p|=rank_t(\p)=\bt$, that is,
$$R_t(\bt)=\{\mbox{error pattern}\ \rho:\ |\rho|=rank_t(\rho)=\bt\}.$$
\end{thm}

The proof of this theorem is the same as that of Theorem \ref{thm_MDS} so long as replace $\dt$ by $\bt$, so the details are omitted.

The following conclusion shows that the required field size for constructing general linear network error correction codes is smaller than that for constructing network MDS codes.
\begin{thm}\label{thm_compare_size}
Let $\beta_t\leq \dt\leq \lfloor\frac{C_t}{2}\rfloor$, then $ |R_t(\beta_t)|\leq |R_t(\dt)|$.
\end{thm}

The proof of Theorem \ref{thm_compare_size} is in Appendix \ref{app}.

\section{The Constructive Algorithm of Linear Network Error Correction Codes}
 From the discussions in the last section, we propose the following Algorithm \ref{algo} for constructing a linear network error correction code with required error correction capability.

\begin{algorithm}[!htp]\label{algo}
\algorithmicrequire{ The single source multicast network $G=(V,E)$, the information rate $w\leq \min_{t\in T}C_t$, and the nonnegative
integers $\bt\leq\dt$ for each $t\in T$.}

\algorithmicensure{ Extended global kernels (forming a linear network error correction code).}

\algorithmicInitialization{
\begin{enumerate}
  \item For each $t\in T$ and each $\p\in R_t(\bt)$, find $(w+\bt)$ channel disjoint paths $\mP_{t,\p}$
  from $In(s)$ or $\p'$ to $t$ satisfying Corollary \ref{coro_path},\;
  \item For each $t\in T$ and each $\p\in R_t(\bt)$, initialize dynamic channel sets
  $CUT_{t,\p}=In(s)\cup\p'\linebreak
  =\{d_1',d_2',\cdots,d_w'\}\cup\{e': e\in\p\},$
      and the extended global encoding kernels
      $\f_e=1_e$ for all imaginary channels $e\in In(s)\cup E'$.\;
\end{enumerate}}
\begin{algorithmic}[1]
\FORALL{node $i\in V$ \rm{(according to the topological order of nodes)}}
    \FORALL{ channel $e\in Out(i)$ \rm{(according to an arbitrary order)}}
        \IF{$e\notin \cup_{t\in T}\cup_{\p\in R_t(\bt)}E_{t,\p}$ }
        \STATE $ \tilde{f}_e=1_e$,
        \STATE all $CUT_{t,\p}$ remain unchanged.
        \ELSIF{$e\in \cup_{t\in T}\cup_{\p\in R_t(\bt)}E_{t,\p}$ }
        \STATE choose $\g_e\in \tilde{\mL}(In(i)\cup\{e'\})\backslash\cup_{t\in T}\cup_{\p\in R_t(\bt):\atop e\in E_{t,\p}}\newline [\mL^{\p}(CUT_{t,\p}\backslash \{e(t,\p)\})+\mL^{\p^c}(In(i)\cup \{e'\})]$,
        \IF{$\g_e(e)=0$}
        \STATE $\f_e=\g_e+1_e$,
        \ELSE
        \STATE $\f_e=\g_e(e)^{-1}\cdot\g_e$.
        \ENDIF
        \STATE For those $CUT_{t,\p}$ satisfying $e\in E_{t,\p}$, update
        $CUT_{t,\p}=\{CUT_{t,\p}\backslash\{e(t,\p)\}\}\cup\{e\}$; and for others, $CUT_{t,\p}$ remain unchanged.
        \ENDIF
\ENDFOR
\ENDFOR
\end{algorithmic}
\caption{The algorithm for constructing a linear network error correction code with error correction capacity
$d_{\min}^{(t)}\geq \bt$ for each $t\in T$.}\label{algo}
\end{algorithm}

\begin{remark}
Similar to the polynomial-time algorithm for constructing linear network codes in {\rm\cite{co-construction}}, our algorithm is a greedy one, too. The verification of Algorithm {\rm\ref{algo}} is from the proof of Theorems {\rm\ref{thm_MDS}} and {\rm\ref{thm_general}}. In particular, if we choose $\bt=\dt$ for all $t\in T$, then, by the proposed algorithm, we can construct a linear network error correction code that meets the refined Singleton bound with equality. That is, we can obtain a linear network error correction MDS code. On the other hand, if we choose $\bt=0$ for each $t\in T$, then this algorithm degenerates into an algorithm for constructing linear network codes.
\end{remark}

Next, we will analyze the time complexity of the proposed algorithm.
First, from \cite{co-construction}, we can determine $R_t(\bt)$ and find $(w+\bt)$ channel disjoint paths satisfying Lemma \ref{lem_path} in time $\mO(\sum_{t\in T}{ |E|\choose \bt}(w+\bt)|E|)$.

Both methods presented by Jaggi \textit{et al.} \cite{co-construction} are used to analyze the time complexity of the main loop.

\begin{itemize}
  \item If we use the method of Testing Linear Independent Quickly \cite[\Rmnum{3},A]{co-construction}, the expected time complexity is at most
      $$\mO\left( |E|\left[\sum_{t\in T}|R_t(\bt)|(w+\bt)(w+\frac{|E|+1}{2})\right]\right).$$
After a simple calculation, the expected time complexity of the algorithm using the method of Testing Linear Independent Quickly is at most
\begin{multline*}
\mO\Big(|E|(w+\bt)\\
         \cdot\left[\sum_{t\in T}{ |E|\choose \bt}+\sum_{t\in T}|R_t(\bt)|(w+\frac{|E|+1}{2})\right] \Big).
\end{multline*}
  \item If we use the method of Deterministic Implementation \cite[\Rmnum{3},B]{co-construction}, the time complexity of the main loop is at most
\begin{multline*}
\mO\Big(|E|(w+\frac{|E|+1}{2})\\
        \cdot\left[(\sum_{t\in T}|R_t(\bt)|)^2+\sum_{t\in T}|R_t(\bt)|(w+\bt)\right]\Big).
\end{multline*}
Therefore, the total time complexity of the algorithm using the method of Deterministic Implementation is at most
\begin{multline*}
\mO\Big( |E|\\
\cdot\Big[(\sum_{t\in T}|R_t(\bt)|)^2(w+\frac{|E|+1}{2})+\sum_{t\in T}{|E|\choose \bt}(w+\bt)\Big]\Big).
\end{multline*}
\end{itemize}

As an example, we will apply Algorithm \ref{algo} to construct a network MDS code for a very simple network $G_1$ shown by Fig. \ref{fig_eg3}.

\begin{eg}\
\begin{figure}[!htb]
\begin{center}
\begin{tikzpicture}[->,>=stealth',shorten >=1pt,auto,node distance=2cm,
                    thick]
  \tikzstyle{every state}=[fill=none,draw=black,text=black,minimum size=7mm]
  \tikzstyle{node}=[circle,fill=none,draw=white, minimum size=0.1mm]
  \node[state]           (s)                     {$s$};
  \node[state]           (i)[below left of=s]    {$i$};
  \node[state]           (t)[below right of=i]   {$t$};
  \node[node]            (s')[above of=s]         {};
  \node[node]            (a)[left of=s',xshift=10mm]{};
  \node[node]            (b)[right of=s',xshift=-10mm]{};
  \node[node]            (c)[above left of=i,xshift=6mm]{};
\path (s)   edge              node[swap]{$e_1$}(i)
            edge[bend left=45]  node{$e_2$}(t)
      (i)   edge              node[swap]{$e_3$}(t)
      (s')  edge[dashed, near start]      node{$d'$}  (s)
      (a)   edge[dashed, near start]      node[swap]{$e_1'$}(s)
      (b)   edge[dashed, near start]      node{$e_2'$}(s)
      (c)   edge[dashed]      node[swap]{$e_3'$}(i);
\end{tikzpicture}
\caption{Network $G_1$.}
\label{fig_eg3}
\end{center}
\end{figure}
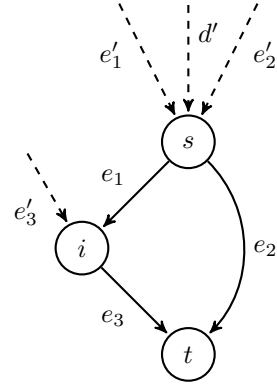
For the network $G_1$ shown by Fig. \ref{fig_eg3}, let the topological order of all nodes be $s\prec i\prec t$, and the topological order of all channels be $e_1\prec e_2\prec e_3$. It is obvious that $C_t=2$. Let $w=1$, and thus $\dt=C_t-w=1$.
Furthermore, we have $R_t(\dt)=R_t(1)=\{\p_1=\{e_1\}, \p_2=\{e_2\}, \p_3=\{e_3\}\}$, and
\begin{align*}
&\mP_{t,\p_1}=\{P_{t,\p_1}^{(\dt)}=(e_1',e_1,e_3),P_{t,\p_1}^{(w)}=(d', e_2) \},\\
&E_{t,\p_1}=\{d',e_1',e_1,e_2,e_3\};\\
&\mP_{t,\p_2}=\{P_{t,\p_2}^{(\dt)}=(e_2',e_2),P_{t,\p_2}^{(w)}=(d', e_1, e_3) \},\\
&E_{t,\p_2}=\{d',e_2',e_1,e_2,e_3\};\\
&\mP_{t,\p_3}=\{P_{t,\p_3}^{(\dt)}=(e_3',e_3),P_{t,\p_3}^{(w)}=(d', e_2) \},\\
&E_{t,\p_3}=\{d',e_3',e_2,e_3\}.
\end{align*}

Let the base field be $\mathbb{Z}_3$.
Initialize the dynamic channel sets $CUT_{t,\p_1}=\{d',e_1'\}$, $CUT_{t,\p_2}=\{d',e_2'\}$, $CUT_{t,\p_3}=\{d',e_3'\}$, and
$$\f_{d'}=\left(\begin{smallmatrix}1\\0\\0\\0\end{smallmatrix}\right), \f_{e_1'}=\left(\begin{smallmatrix}0\\1\\0\\0\end{smallmatrix}\right), \f_{e_2'}=\left(\begin{smallmatrix}0\\0\\1\\0\end{smallmatrix}\right), \f_{e_3'}=\left(\begin{smallmatrix}0\\0\\0\\1\end{smallmatrix}\right),$$
which leads to $\dim(\tilde{\mL}^{\p_i}(CUT_{t,\p_i}))=2,\ (i=1,2,3)$.

For the channel $e_1\in Out(s)$, $e_1\in E_{t,\p_1}\cap E_{t,\p_2}$ and
\begin{multline*}
\tilde{\mL}(\{d',e_1'\})\backslash[\mL^{\p_1}(\{d'\})+\mL^{\p_1^c}(\{d',e_1'\})]\\
\cup[\mL^{\p_2}(\{e_2'\})+\mL'^{\p_2^c}(\{d',e_1'\})]\\
=\left\langle
\left(\begin{smallmatrix}1\\0\\0\\0\end{smallmatrix}\right),
\left(\begin{smallmatrix}0\\1\\0\\0\end{smallmatrix}\right)
\right\rangle
\Big\backslash
\left\langle
\left(\begin{smallmatrix}1\\0\\0\\0\end{smallmatrix}\right)
\right\rangle
\cup
\left\langle
\left(\begin{smallmatrix}0\\0\\1\\0\end{smallmatrix}\right)
\right\rangle
+
\left\langle
\left(\begin{smallmatrix}0\\1\\0\\0\end{smallmatrix}\right)
\right\rangle.\hfill
\end{multline*}
So we choose $\g_{e_1}=\left(\begin{smallmatrix}1\\1\\0\\0\end{smallmatrix}\right)$ because of
\begin{multline*}
\g_{e_1}\in \tilde{\mL}(\{d',e_1'\})\backslash[\mL^{\p_1}(\{d'\})+\mL^{\p_1^c}(\{d',e_1'\})]\\
\cup[\mL^{\p_2}(\{e_2'\})+\mL'^{\p_2^c}(\{d',e_1'\})].
\end{multline*}
And $\f_{e_1}=\g_{e_1}$, since $\g_{e_1}(e_1)=1$. Then update $CUT_{t,\p_1}=\{d',e_1\}$, $CUT_{t,\p_2}=\{e_1,e_2'\}$, and $CUT_{t,\p_3}$ remains unchanged.

For the channel $e_2\in Out(s)$, $e_2\in E_{t,\p_1}\cap E_{t,\p_2}\cap E_{t,\p_3}$ and
\begin{multline*}
\tilde{\mL}(\{d',e_2'\})\backslash[\mL^{\p_1}(\{e_1\})+\mL^{\p_1^c}(\{d',e_2'\})]\\
\cup [\mL^{\p_2}(\{e_1\})+\mL^{\p_2^c}(\{d',e_2'\})]\cup[\mL^{\p_3}(\{e_3'\})+\mL^{\p_3^c}(\{d',e_2'\})]\\
=\left\langle \left(\begin{smallmatrix}1\\0\\0\\0\end{smallmatrix}\right),
\left(\begin{smallmatrix}0\\0\\1\\0\end{smallmatrix}\right)
\right\rangle
\Big\backslash \left\langle
\left(\begin{smallmatrix}1\\1\\0\\0\end{smallmatrix}\right)
\right\rangle+\left\langle
\left(\begin{smallmatrix}0\\0\\1\\0\end{smallmatrix}\right)
\right\rangle\hfill\\
\cup
\left\langle
\left(\begin{smallmatrix}1\\0\\0\\0\end{smallmatrix}\right)
\right\rangle
\cup
\left\langle
\left(\begin{smallmatrix}0\\0\\0\\1\end{smallmatrix}\right)
\right\rangle
+
\left\langle
\left(\begin{smallmatrix}0\\0\\1\\0\end{smallmatrix}\right)
\right\rangle.
\end{multline*}
We choose $\g_{e_2}=\left(\begin{smallmatrix}1\\0\\1\\0\end{smallmatrix}\right)$, since
\begin{multline*}
\g_{e_2}\in \tilde{\mL}(\{d',e_2'\})\backslash[\mL^{\p_1}(\{e_1\})+\mL^{\p_1^c}(\{d',e_2'\})]\\
\cup [\mL^{\p_2}(\{e_1\})+\mL^{\p_2^c}(\{d',e_2'\})]\cup[\mL^{\p_3}(\{e_3'\})+\mL^{\p_3^c}(\{d',e_2'\})],
\end{multline*}
which, together with $\g_{e_2}(e_2)=1$, shows that $\f_{e_2}=\g_{e_2}$.
Then, update $CUT_{t,\p_1}=\{e_2,e_1\}$, $CUT_{t,\p_2}=\{e_1,e_2\}$, and $CUT_{t,\p_3}=\{e_2, e_3'\}$.

For the channel $e_3\in Out(i)$, $e_3\in E_{t,\p_1}\cap E_{t,\p_2}\cap E_{t,\p_3}$ and
\begin{multline*}
\tilde{\mL}(\{e_1,e_3'\})\backslash[\mL^{\p_1}(\{e_2\})+\mL^{\p_1^c}(\{e_1,e_3'\})]\\
\cup[\mL^{\p_2}(\{e_2\})+\mL^{\p_2^c}(\{e_1,e_3'\})]\cup[\mL^{\p_3}(\{e_2\})+\mL^{\p_3^c}(\{e_1,e_3'\})]\\
=\left\langle
\left(\begin{smallmatrix}1\\1\\0\\0\end{smallmatrix}\right),
\left(\begin{smallmatrix}0\\0\\0\\1\end{smallmatrix}\right)
\right\rangle
\Big\backslash
\left\langle
\left(\begin{smallmatrix}1\\0\\0\\0\end{smallmatrix}\right)
\right\rangle
+\left\langle
\left(\begin{smallmatrix}0\\0\\0\\1\end{smallmatrix}\right)
\right\rangle \hfill\\
\cup\left\langle
\left(\begin{smallmatrix}1\\0\\1\\0\end{smallmatrix}\right)
\right\rangle+
\left\langle \left(\begin{smallmatrix}0\\1\\0\\0\end{smallmatrix}\right),
\left(\begin{smallmatrix}0\\0\\0\\1\end{smallmatrix}\right)
\right\rangle\cup\left\langle \left(\begin{smallmatrix}1\\0\\0\\0\end{smallmatrix}\right)
\right\rangle
+\left\langle
\left(\begin{smallmatrix}0\\1\\0\\0\end{smallmatrix}\right)
\right\rangle.
\end{multline*}
We select $\g_{e_3}=\left(\begin{smallmatrix}1\\1\\0\\1\end{smallmatrix}\right)$ satisfying
\begin{multline*}
\g_{e_3}\in \tilde{\mL}(\{e_1,e_3'\})\backslash[\mL^{\p_1}(\{e_2\})+\mL^{\p_1^c}(\{e_1,e_3'\})]\\
\cup[\mL^{\p_2}(\{e_2\})+\mL^{\p_2^c}(\{e_1,e_3'\})]\cup[\mL^{\p_3}(\{e_2\})+\mL^{\p_3^c}(\{e_1,e_3'\})].
\end{multline*}
It follows that $\f_{e_3}=\g_{e_3}$ from $\g_{e_3}(e_3)=1$, and update $CUT_{t,\p_1}=CUT_{t,\p_2}=CUT_{t,\p_3}=\{e_2, e_3\}\subseteq In(t)$.

The decoding matrix at $t$ is $\tilde{F}_t=(\f_{e_2}\ \f_{e_3})=\left(\begin{smallmatrix}1&1\\0&1\\1&0\\0&1\end{smallmatrix}\right)$. It is easy to check that $\Phi(t)\cap\Delta(t,\p_i)=\{\underline{0}\}$ for $i=1,2,3$.
Further, let $\p=\{e_1,e_2\}$. Then $rank_t(\p)=2$ and $\Phi(t)\cap\Delta(t,\p)\neq\{\underline{0}\}$, which means $d_{\min}^{(t)}=2=\dt+1$.
That is, $\{\f_{e_1}, \f_{e_2}, \f_{e_3}\}$ forms a global description of a linear network error correction MDS code for the network $G_1$.
\end{eg}

\section{Random Linear Network Error Correction Coding}
Random network coding was originally proposed in \cite{Ho-etc-random}. When a node (maybe the source node $s$) receives the messages from its all incoming channels, for each outgoing channel, it selects the encoding coefficients uniformly at random over the base field $\mF$, uses them to encode the messages
and transmits the encoded messages over the outgoing channel. In other words, the local encoding coefficients $k_{d,e}$ are independently, uniformly distributed random variables on the base field $\mF$. The performance analysis of random linear network coding is very important in theory and applications.
In this section, we will investigate the error correction capability of random linear network coding. We first consider random linear network error correction MDS codes. Before the discussion, we give the following definitions.

\begin{defn}
Let $G$ be a single source multicast network, $\mathbf{C}$ be a random linear network error correction code on $G$, and $d_{\min}^{(t)}$ be the minimum distance at sink node $t$ of $\mathbf{C}$.
\begin{itemize}
  \item $P_{ec}(t)\triangleq Pr(\{\dim(\Phi(t))<w\}\cup\{ d_{\min}^{(t)}< \dt+1\})$ is called the failure probability of random linear network error correction MDS coding for sink node $t$.
  \item $P_{ec}\triangleq Pr(\{\mathbf{C}\mbox{ is not regular}\}\cup\{\exists\ t\in T\mbox{such that\:} d_{\min}^{(t)}< \dt+1\})$ is called the failure probability of random linear network error correction MDS coding for network $G$, that is the probability that network MDS codes are not constructed by the random method.
\end{itemize}
\end{defn}

In order to evaluate these two failure probabilities, the following lemma is useful.

\begin{lemma}[{\cite[Lemma 1]{Guang-Fu-random},\cite{zhang-course}}]{\label{lem_bound}}
Let $\mathcal{L}$ be an $n$ dimensional linear space over a finite field $\mathcal{F}$,
$\mathcal{L}_0,\ \mathcal{L}_1$ be two subspaces of $\mathcal{L}$
of dimensions $k_0,\ k_1$, respectively, and
$\langle\mathcal{L}_0\cup\mathcal{L}_1\rangle=\mathcal{L}$. Let
$l_1,\ l_2,\ \cdots,\ l_{m}$ $(m=n-k_0)$ be $m$ independently and
uniformly distributed random vectors taking values in
$\mathcal{L}_1$. Then
$$Pr(\dim(\langle \mathcal{L}_0 \cup \{l_1,\
l_2,\ \cdots,\ l_{m}\}\rangle)=n)=\prod_{i=1}^{m}\left(
1-\frac{1}{\mathcal{|F|}^i}\right). $$
\end{lemma}

\begin{thm}\label{thm_random_MDS}
Let $G$ be a single source multicast network, and $w\leq \min_{t\in T}C_t$. Using random method to construct a linear network error correction MDS code, then
\begin{itemize}
  \item for each $t\in T$, the failure probability of random linear network error correction MDS coding for $t$ satisfies
$$P_{ec}(t)<1-\left( 1-\frac{|R_t(\dt)|}{|\mF|-1} \right)^{|J|+1};$$
  \item the failure probability of random linear network error correction MDS coding for the network $G$ satisfies
$$P_{ec}<1-\left( 1-\frac{\sum_{t\in T}|R_t(\dt)|}{|\mF|-1} \right)^{|J|+1},$$
where $J$ is the set of the internal nodes in $G$.
\end{itemize}
\end{thm}
\begin{IEEEproof}
For the single source multicast network $G=(V,E)$, $s$ is the single source node, $T$ is the set of the sink nodes, $J=V-\{s\}-T$ is the set of the internal nodes, and $E$ is the set of all channels. Let $\tilde{G}=(\tilde{V},\tilde{E})$ be the extended network of $G$.

For each sink node $t\in T$ and each error pattern $\p\in R_t(\dt)$, Corollary \ref{coro_path} implies that there are $(w+\dt)$ channel disjoint paths from either $In(s)$ or $\p'$ to $t$ satisfying the properties that (1) there exist exactly $\dt$ channel disjoint paths from $\p'$ to $t$, and $w$ channel disjoint paths from $In(s)$ to $t$; (2) each of these $\dt$ paths from $\p'$ to $t$ starts with a channel $e'\in\p'$ and passes through the corresponding channel $e\in \p$. Denote by $\mP_{t,\p}$ the set of $(w+\dt)$ channel disjoint paths satisfying these properties and $E_{t,\p}$ denotes the set of all channels in $\mP_{t,\p}$.

Note that the event ``$\{\dim(\Phi(t))=w\}\cap\{ d_{\min}^{(t)}= \dt+1\}$'' is equivalent to the event ``$\{\dim(\Phi(t))=w\}\cap\{\forall\ \p\in R_t(\dt): \Phi(t)\cap \Delta(t,\p)=\{\underline{0}\}\}$'', and furthermore, the event ``$\forall\ \p\in R_t(\dt): \Rank(\tilde{F}_t^{\p})=w+\dt$'' implies the event ``$\{\dim(\Phi(t))=w\}\cap\{\forall\ \p\in R_t(\dt): \Phi(t)\cap \Delta(t,\p)=\{\underline{0}\}\}$''. Thus, we consider the following probability:
$$Pr(\cap_{\p\in R_t(\dt)}\Rank(\tilde{F}_t^{\p})=w+\dt).$$

For the network $G$, let an ancestral order of nodes be
$$s\prec i_1\prec i_2 \prec \cdots \prec i_{|J|} \prec T .$$
During our discussion, we use the concept of cuts of the paths similar to the dynamic set $CUT_{t,\p}$ as mentioned above. The first cut is $CUT_{t,\p,0}=In(s)\cup\{e': e\in \p\}$, i.e., the $w$ imaginary message channels $d_1',d_2',\cdots,d_w'$ and imaginary error channels corresponding to the channels in $\p$. At node $s$, the next $CUT_{t,\p,1}$ is formed from $CUT_{t,\p,0}$ by replacing those channels in $\{In(s)\cup \{e': e\in Out(s)\}\}\cap CUT_{t,\p,0}$ by their respective next channels in the paths. These new channels are in $Out(s)\cap E_{t,\p}$. Other channels remain the same as in $CUT_{t,\p,0}$. At node $i_1$, the next cut $CUT_{t,\p,2}$ is formed from $CUT_{t,\p,1}$ by replacing those channels in $\{In(i_1)\cup \{e': e\in Out(i_1)\}\}\cap CUT_{t,\p,1}$ by their respective next channels in the paths. These new channels are in $Out(i_1)\cap E_{t,\p}$. Other channels remain the same as in $CUT_{t,\p,1}$. Subsequently, once $CUT_{t,\p,k}$ is defined, $CUT_{t,\p,k+1}$ is formed from $CUT_{t,\p,k}$ by the same method. By induction, all cuts $CUT_{t,\p,k}$ for $t\in T,\ \p\in R_t(\dt)$, and $k=0,1,2,\cdots, |J|+1$ can be defined.
Moreover, for each $CUT_{t,\p,k}$, we divide $CUT_{t,\p,k}$ into two disjoint parts $CUT_{t,\p,k}^{in}$ and $CUT_{t,\p,k}^{out}$ as follows:
\begin{align*}
CUT_{t,\p,k}^{in}&=\{e: e\in CUT_{t,\p,k}\cap In(i_{k})\},\\
CUT_{t,\p,k}^{out}&=\{e: e\in CUT_{t,\p,k}\backslash CUT_{t,\p,k}^{in}\}.
\end{align*}

Define $(w+\dt)\times(w+\dt)$ matrix $\tilde{F}_{t}^{\p(k)}=(\f_{e}^{\p}: e\in CUT_{t,\p,k})$ for $k=0,1,\cdots,|J|+1$.
If $\Rank(\tilde{F}_{t}^{\p(k)})<w+\dt$, we call that we have a failure at $CUT_{t,\p,k}$. Let $\Gamma_k^{(t,\p)}$ represent the event ``$\Rank(\tilde{F}_{t}^{\p(k)})=w+\dt$''. Furthermore, let $|J|=m$, and note that $\tilde{F}_{t}^{\p(m+1)}$ is a submatrix of $\tilde{F}_t^{\p}$. It follows that the event ``$\forall \p\in R_t(\dt), \Rank(\tilde{F}_{t}^{\p(m+1)})=w+\dt$'' implies the event ``$\forall \p\in R_t(\dt), \Rank(\tilde{F}_{t}^{\p})=w+\dt$''. Therefore,
\begin{align*}
&1-P_{ec}(t)\\
=&Pr(\{\dim(\Phi(t))=w\}\cap\{d_{\min}^{(t)}=\dt+1\})\\
=&Pr(\{\dim(\Phi(t))=w\}\cap \{\cap_{\p\in R_t(\dt)} \Phi(t)\cap \Delta(t,\p)=\{\underline{0}\}\})\\
\geq&Pr(\cap_{\p\in R_t(\dt)}\Rank(\tilde{F}_t^{\p})=w+\dt)\\
\geq& Pr(\cap_{\p\in R_t(\dt)}\Gamma_{m+1}^{(t,\p)}).
\end{align*}
Consequently,
\begin{align}
&Pr(\cap_{\p\in R_t(\dt)}\Gamma_{m+1}^{(t,\p)})\nonumber\\
\geq&Pr(\cap_{\p\in R_t(\dt)}\Gamma_{m+1}^{(t,\p)},\cap_{\p\in R_t(\dt)}\Gamma_{m}^{(t,\p)},\cdots,\cap_{\p\in R_t(\dt)}\Gamma_{0}^{(t,\p)})\nonumber\\
\geq&Pr(\cap_{\p\in R_t(\dt)}\Gamma_{m+1}^{(t,\p)}|\cap_{\p\in R_t(\dt)}\Gamma_{m}^{(t,\p)})\cdots \nonumber\\ &Pr(\cap_{\p\in R_t(\dt)}\Gamma_{1}^{(t,\p)}|\cap_{\p\in R_t(\dt)}\Gamma_{0}^{(t,\p)})Pr(\cap_{\p\in R_t(\dt)}\Gamma_{0}^{(t,\p)})\nonumber\\
=&\prod_{k=0}^{m}Pr(\cap_{\p\in R_t(\dt)}\Gamma_{k+1}^{(t,\p)}|\cap_{\p\in R_t(\dt)}\Gamma_{k}^{(t,\p)}), \label{Pr}
\end{align}
where (\ref{Pr}) follows from
\begin{align*}
&Pr(\cap_{\p\in R_t(\dt)}\Gamma_{0}^{(t,\p)})\\
=&Pr(\cap_{\p\in R_t(\dt)}\Rank((\f_e^{\p}: e\in In(s)\cup \p'))=w+\dt)\\
=&Pr(\Rank(I_{w+\dt})=w+\dt)\equiv 1.
\end{align*}

For each channel $e\in E$, let $e\in Out(i_{k})$. Let $\g_e$ be an independently and uniformly distributed random vector taking values in $\tilde{\mL}(In(i_{k}))$. In other words, if $In(i_{k})=\{d_1,d_2,\cdots,d_l\}$, then
$$\g_e=k_{d_1,e}\f_{d_1}+k_{d_2,e}\f_{d_2}+\cdots+k_{d_l,e}\f_{d_l},$$
where $k_{d_j,e}\ (j=1,2,\cdots,l)$ are independently and uniformly distributed random variables taking values in the base field $\mF$. It follows that $\g_e^{\p}=k_{d_1,e}\f_{d_1}^{\p}+k_{d_2,e}\f_{d_2}^{\p}+\cdots+k_{d_l,e}\f_{d_l}^{\p}$ is also an independently and uniformly distributed random vector taking values in $\tilde{\mL}^{\p}(In(i_{k}))$. We always define $\f_e=\g_e+1_e$.
Therefore, for all $e\in E_{t,\p}\cap Out(i_{k})$ with $e(t,\p)\in CUT_{t,\p,k}^{in}$, i.e., $e\notin \p$, it is shown that $\f_e^{\p}=\g_{e}^{\p}$ because of $e\notin \p$. Thus, $\f_e^{\p}$ is an independently and uniformly distributed random vector taking values in $\tilde{\mL}^{\p}(In(i_{k}))$. Otherwise $e\in E_{t,\p}\cap Out(i_{k})$ with $e(t,\p)\in CUT_{t,\p,k}^{out}$, that is, $e(t,\p)=e'$, then, $\f_e^{\p}$ and $\{\f_{d}^{\p}: d\in CUT_{t,\p,k}\backslash e(t,\p)\}$ are always linearly independent, since $\f_e^{\p}(e)=1$ and $\f_{d}^{\p}(e)=0$ for all $d\in CUT_{t,\p,k}\backslash e(t,\p)$.

Applying Lemma \ref{lem_bound}, we derive
\begin{align*}
&Pr(\Gamma_{k+1}^{(t,\p)}|\Gamma_k^{(t,\p)})
=\prod_{i=1}^{|CUT_{t,\p,k}^{in}|}\left( 1-\frac{1}{|\mF|^i} \right)\\
\geq&\prod_{i=1}^{w+\dt}\left( 1-\frac{1}{|\mF|^i}\right)>1-\sum_{i=1}^{w+\dt}\frac{1}{|\mF|^i}\\
>&1-\sum_{i=1}^{\infty}\frac{1}{|\mF|^i}=1-\frac{1}{|\mF|-1}.
\end{align*}
Consequently, for each $k\ (0\leq k\leq m)$, one has
\begin{align*}
&Pr(\cap_{\p\in R_t(\dt)}\Gamma_{k+1}^{(t,\p)}|\cap_{\p\in R_t(\dt)}\Gamma_{k}^{(t,\p)})\\
=&1-Pr(\cup_{\p\in R_t(\dt)}{\Gamma_{k+1}^{(t,\p)}}^c|\cap_{\p\in R_t(\dt)}\Gamma_{k}^{(t,\p)})\\
\geq&1-\sum_{\p\in R_t(\dt)}Pr({\Gamma_{k+1}^{(t,\p)}}^c|\cap_{\p\in R_t(\dt)}\Gamma_{k}^{(t,\p)})\\
=&1-\sum_{\p\in R_t(\dt)}Pr({\Gamma_{k+1}^{(t,\p)}}^c|\Gamma_{k}^{(t,\p)})\\
>&1-\sum_{\p\in R_t(\dt)}\frac{1}{|\mF|-1}\\
=&1-\frac{|R_t(\dt)|}{|\mF|-1}.
\end{align*}

Combining the above inequalities, we have
\begin{align*}
1-P_{ec}(t)\geq&\prod_{k=0}^{m}Pr(\cap_{\p\in R_t(\dt)}\Gamma_{k+1}^{(t,\p)}|\cap_{\p\in R_t(\dt)}\Gamma_{k}^{(t,\p)})\\
>&\left( 1-\frac{|R_t(\dt)|}{|\mF|-1} \right)^{m+1}.
\end{align*}
That is,
$$P_{ec}(t)<1-\left( 1-\frac{|R_t(\dt)|}{|\mF|-1} \right)^{m+1}.$$

Next,
\begin{align}
&1-P_{ec}\geq Pr(\cap_{t\in T}\cap_{\p\in R_t(\dt)}Rank(\tilde{F}_t^{\p})=w+\dt)\nonumber\\
\geq&Pr(\cap_{t\in T}\cap_{\p\in R_t(\dt)}\Gamma_{m+1}^{(t,\p)},\cap_{t\in T}\cap_{\p\in R_t(\dt)}\Gamma_{m }^{(t,\p)}, \cdots, \nonumber\\
&\cap_{t\in T}\cap_{\p\in R_t(\dt)}\Gamma_{0}^{(t,\p)})\nonumber\\
\geq&Pr(\cap_{t\in T}\cap_{\p\in R_t(\dt)}\Gamma_{m+1}^{(t,\p)}|\cap_{t\in T}\cap_{\p\in R_t(\dt)}\Gamma_{m}^{(t,\p)})\nonumber\\
&\cdot Pr(\cap_{t\in T}\cap_{\p\in R_t(\dt)}\Gamma_{m}^{(t,\p)}|\cap_{t\in T}\cap_{\p\in R_t(\dt)}\Gamma_{m-1}^{(t,\p)})\cdots\nonumber\\
&\cdot Pr(\cap_{t\in T}\cap_{\p\in R_t(\dt)}\Gamma_{1}^{(t,\p)}|\cap_{t\in T}\cap_{\p\in R_t(\dt)}\Gamma_{0}^{(t,\p)})\label{PrMDS1}\\
=&\prod_{k=0}^{m}Pr(\cap_{t\in T}\cap_{\p\in R_t(\dt)}\Gamma_{k+1}^{(t,\p)}|\cap_{t\in T}\cap_{\p\in R_t(\dt)}\Gamma_{k}^{(t,\p)}),\nonumber
\end{align}
where (\ref{PrMDS1}) follows from $Pr(\cap_{t\in T}\cap_{\p\in R_t(\dt)}\Gamma_{0}^{(t,\p)})\equiv 1$.

Furthermore, for each $k\ (0\leq k \leq m)$,
\begin{align}
&Pr(\cap_{t\in T}\cap_{\p\in R_t(\dt)}\Gamma_{k+1}^{(t,\p)}|\cap_{t\in T}\cap_{\p\in R_t(\dt)}\Gamma_{k}^{(t,\p)})\nonumber\\
=&1-Pr(\cup_{t\in T}\cup_{\p\in R_t(\dt)}{\Gamma_{k+1}^{(t,\p)}}^c|\cap_{t\in T}\cap_{\p\in R_t(\dt)}\Gamma_{k}^{(t,\p)})\nonumber\\
\geq&1-\sum_{t\in T}\sum_{\p\in R_t(\dt)}Pr({\Gamma_{k+1}^{(t,\p)}}^c|\Gamma_{k}^{(t,\p)})\nonumber\\
=&1-\sum_{t\in T}\sum_{\p\in R_t(\dt)}[1-Pr({\Gamma_{k+1}^{(t,\p)}}|\Gamma_{k}^{(t,\p)})]\nonumber\\
>&1-\sum_{t\in T}\sum_{\p\in R_t(\dt)}\frac{1}{|\mF|-1}\nonumber\\
=&1-\frac{\sum_{t\in T}|R_t(\dt)|}{|\mF|-1}.\label{PrMDS2}
\end{align}
Combining the inequalities (\ref{PrMDS1}) and (\ref{PrMDS2}), we have
$$1-P_{ec}>\left( 1-\frac{\sum_{t\in T}|R_t(\dt)|}{|\mF|-1} \right)^{m+1},$$
that is,
$$P_{ec}<1-\left( 1-\frac{\sum_{t\in T}|R_t(\dt)|}{|\mF|-1} \right)^{m+1}.$$
The proof is completed.
\end{IEEEproof}

Applying Lemma \ref{lem_field_size} to Theorem \ref{thm_random_MDS}, we derive the following corollary.
\begin{cor}
The failure probability $P_{ec}(t)$ of random linear network error correction MDS coding for each $t\in T$ satisfies
$$P_{ec}(t)<1-\left( 1-\frac{{|E_t|\choose \dt}}{|\mF|-1} \right)^{|J|+1}
\leq 1-\left( 1-\frac{{|E|\choose \dt}}{|\mF|-1} \right)^{|J|+1}.$$
The failure probability $P_{ec}$ of random linear network error correction MDS coding for the network $G$ satisfies
\begin{align*}
P_{ec}<&1-\left( 1-\frac{\sum_{t\in T}{|E_t|\choose \dt}}{|\mF|-1} \right)^{|J|+1}\\
\leq & 1-\left( 1-\frac{\sum_{t\in T}{|E|\choose \dt}}{|\mF|-1} \right)^{|J|+1}.
\end{align*}
\end{cor}

However, in practice, we sometimes need general linear network error correction codes instead of the network MDS codes. That is, we only need the codes satisfying that its minimum distance $d_{\min}^{(t)}\geq \bt$, where $\bt \leq \dt$ is a nonnegative integer. The part of reason is that usually the field size required by general linear network error correction codes is smaller than that of network MDS codes. Hence, we should also discuss the random method for the general linear network error correction codes. Similarly, we define the failure probabilities for random linear network error correction codes as follows.
\begin{defn}
Let $G$ be a single source multicast network, $\mathbf{C}$ be a random linear network error correction code on $G$, and $d_{\min}^{(t)}$ be the minimum distance at sink node $t$. Define that
\begin{itemize}
  \item $P_{ec}(t,\bt)\triangleq Pr(\{ \dim(\Phi(t))<w \}\cup\{ d_{\min}^{(t)}< \bt+1\})$, that is the probability that the code $\mathbf{C}$ cannot either be decoded or satisfy that the error correction capacity $d_{\min}^{(t)}\geq \bt+1$ at the sink node $t$;
  \item $P_{ec}(\bt)\triangleq Pr(\{\ \mathbf{C}\mbox{ is not regular }\}\cup\{\exists\ t\in T \mbox{ such that }
      \\d_{\min}^{(t)}<\bt+1\})$, that is the probability that the regular linear network error correction codes with $d_{\min}^{(t)}\geq \bt+1$ cannot be constructed by the random method.
\end{itemize}
\end{defn}

Using the similar method to prove Theorem \ref{thm_random_MDS}, and combining it with the method to prove the random linear network coding with proper redundancy \cite[Theorem 2]{zhang-random}, we can get the following results.

\begin{thm}\label{thm_random_general}
Let $G$ be a single source multicast network, the minimum cut capacity for sink node $t\in T$ be $C_t$ and the information rate be $w$ symbols per unit time satisfying $w\leq \min_{t\in T}C_t$. Using random method to construct a linear network error correction code, then
\begin{itemize}
  \item for each $t\in T$ and $\bt\leq \dt$,
$$P_{ec}(t,\bt)\leq \frac{|R_t(\bt)|{\dt-\bt+|J|+1\choose |J|}}{(|\mF|-1)^{\dt-\bt+1}};$$
  \item for the network $G$,
$$P_{ec}(\bt)\leq \sum_{t\in T}\frac{|R_t(\bt)|{\dt-\bt+|J|+1\choose |J|}}{(|\mF|-1)^{\dt-\bt+1}}.$$
\end{itemize}
\end{thm}
\begin{remark}
Both Theorems \ref{thm_random_MDS} and \ref{thm_random_general} above imply that these failure probabilities can become arbitrarily small when the size of the base field $\mF$ is sufficiently large.
\end{remark}

Balli, Yan, and Zhang \cite{zhang-random} used $D_{\min}^{(t)}$ to denote the minimum distance of random linear network error correction code at a sink node $t\in T$. Obviously, the refined Singleton bound tells us that $D_{\min}^{(t)}$ takes values in $\{0,1,2,\cdots,\dt+1\}$. Furthermore, they studied the probability mass function of $D_{\min}^{(t)}$. For a code with the minimum distance $d_{\min}^{(t)}$ at sink node $t$, $\dt+1-d_{\min}^{(t)}$ is called the degradation of the code at $t$. Then they presented the following conclusions.

\begin{prop}[{\cite[Theorem 4]{zhang-random}}]
For single source multicast over an acyclic
network $G$, let the minimum cut capacity for sink node
$t\in T$ be $C_t$, let the information rate be $w$ symbols per unit time,
let $\dt=C_t-w$ be the redundancy of the code for the sink node $t\in T$.
For a given $d\geq0$, the linear random network code satisfies:
$$Pr(D_{\min}^{(t)}<\dt+1-d)\leq \frac{{|E|\choose \dt-d}{d+|J|+1\choose |J|}}{(|\mF|-1)^{d+1}}.$$
Furthermore, the probability that random linear network code
has minimum distance at least $\dt+1-d$ at all sinks $t\in T$ is
lower bounded by,
$$Pr(D_{\min}^{(t)}\geq\dt+1-d, \forall\ t\in T)\geq 1-\sum_{t\in T}\frac{{|E|\choose \dt-d}{d+|J|+1\choose |J|}}{(|\mF|-1)^{d+1}}.$$
\end{prop}

This proposition can lead to an upper bound on the field size required for the existence of linear network error correction codes with degradation at most $d$.

\begin{prop}[{\cite[Corollary 1]{zhang-random}}]
If the field size satisfies the following condition:
$$|\mF|\geq 2+\left( \sum_{t\in T}{|E|\choose \dt-d}{d+|J|+1\choose |J|} \right)^{\frac{1}{d+1}},$$
then there exists a code having degradation at most $d$ at all
sinks $t\in T$.
\end{prop}

In the same way, applying Theorem \ref{thm_random_general}, we can also get a probability mass function of $D_{\min}^{(t)}$.

\begin{cor}
For a single source multicast network $G=(V,E)$, let the minimum cut capacity for sink node $t\in T$ be $C_t$, the information rate be $w$ symbols per unit time satisfying $w\leq \min_{t\in T}C_t$, and $\dt=C_t-w$ be the redundancy of the code for sink $t\in T$. For a given $d\geq0$, the random linear network error correction codes satisfy:
$$Pr(D_{\min}^{(t)}<\dt+1-d)\leq \frac{|R_t(\dt-d)|{d+|J|+1\choose |J|}}{(|\mF|-1)^{d+1}},$$
and
\begin{align*}
&Pr(D_{\min}^{(t)}\geq\dt+1-d, \forall\ t\in T)\\
\geq& 1-\sum_{t\in T}\frac{|R_t(\dt-d)|{d+|J|+1\choose |J|}}{{(|\mF|-1)^{d+1}}}.
\end{align*}
\end{cor}

This corollary also leads to an upper bound on the field size required for the existence of linear network error correction codes with degradation at most $d$. On the other hand, Theorem \ref{thm_general} shows that the required field size satisfies $|\mF|\geq \sum_{t\in T}|R(\bt)|$. Therefore, we derive the following result.
\begin{cor}\label{cor_field_size}
If the size of the base field $\mF$ satisfies the following condition:
\begin{multline*}
|\mF|\geq \min\Big\{ \sum_{t\in T}|R_t(\dt-d)|,\\
 2+\left[ \sum_{t\in T}|R(\dt-d)|{d+|J|+1 \choose |J|} \right]^{\frac{1}{d+1}} \Big\},
\end{multline*}
then there exists a regular linear network error correction code having degradation at most $d$ at all
sink nodes $t\in T$.
\end{cor}

When $d=0$, it is readily seen that
\begin{align*}
&\sum_{t\in T}|R_t(\dt-d)|\\
=&\sum_{t\in T}|R_t(\dt)|\\
<&2+(|J|+1)\sum_{t\in T}|R_t(\dt)|\\
=&2+\left[ \sum_{t\in T}|R_t(\dt-d)|{d+|J|+1 \choose |J|} \right]^{\frac{1}{d+1}}.
\end{align*}
This means that, for network MDS codes, Corollary \ref{cor_field_size} cannot give a smaller field size required.
But, for $d\geq 1$, the size bounds $2+\left[ \sum_{t\in T}|R_t(\dt-d)|{d+|J|+1 \choose |J|} \right]^{\frac{1}{d+1}}$ and $\sum_{t\in T}|R_t(\dt-d)|$ have no deterministic relations. We will illustrate this point through the following example.
\begin{eg}
For network $G_2$ shown by Fig. \ref{fig_eg1} below, let $w=2$. Then $\dt=C_t-w=2$.
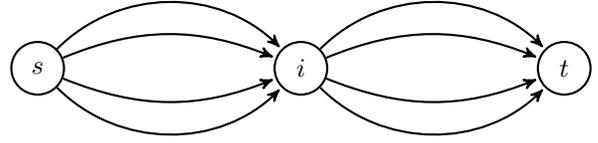
\begin{figure}[!htb]
\begin{center}
\begin{tikzpicture}[->,>=stealth',shorten >=1pt,auto,node distance=3.5cm,
                    thick]
  \tikzstyle{every state}=[fill=none,draw=black,text=black,minimum size=7mm]
  \tikzstyle{place}=[fill=none,draw=white,minimum size=0.1mm]
  \node[state]         (s)               {$s$};
  \node[state]         (i)[right of=s]   {$i$};
  \node[state]         (t)[right of=i]   {$t$};
  \path (s) edge [bend left=22.5]       node {} (i)
            edge [bend left=45]  node [swap]{} (i)
            edge [bend right=22.5]   node {} (i)
            edge [bend right=45]  node[swap] {} (i)
        (i) edge [bend left=45]       node {} (t)
            edge [bend left=22.5]  node [swap]{} (t)
            edge [bend right=22.5]   node {} (t)
            edge [bend right=45]  node[swap] {} (t);
\end{tikzpicture}
\caption{Network $G_2$ with $|T|=1$, $|J|=1$, $C_t=4$.}
\label{fig_eg1}
\end{center}
\end{figure}
\begin{itemize}
  \item In the case $d=0$, it is clear that
  \begin{align*}
  &2+\left[ \sum_{t\in T}|R_t(\dt-d)|{d+|J|+1 \choose |J|} \right]^{\frac{1}{d+1}}\\
  =&2+2|R_t(2)|>|R_t(2)|.
  \end{align*}
  \item In the case $d=1$, a simple calculation gives $$\sum_{t\in T}|R_t(\dt-d)|=|R_t(1)|=8,$$
  and
  \begin{align*}
  &2+\left[ \sum_{t\in T}|R_t(\dt-d)|{d+|J|+1 \choose |J|} \right]^{\frac{1}{d+1}}\\
  =&2+\sqrt{24}<2+5=7.
  \end{align*}
  This shows that in this case
  \begin{align*}
  &2+\left[ \sum_{t\in T}|R_t(\dt-d)|{d+|J|+1 \choose |J|} \right]^{\frac{1}{d+1}}\\
  <&\sum_{t\in T}|R_t(\dt-d)|.
  \end{align*}
\end{itemize}
Nevertheless, for the network $G_3$ shown by Fig. \ref{fig_eg2}, let $w=2$, which shows $\dt=C_t-w=2$.
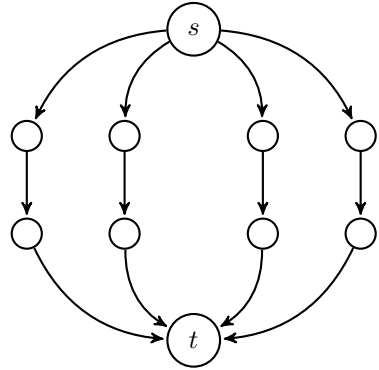
\begin{figure}[!htb]
\begin{center}
\begin{tikzpicture}[->,>=stealth',shorten >=1pt,auto,node distance=1.3cm,
                    thick]
  \tikzstyle{every state}=[fill=none,draw=black,text=black,minimum size=7mm]
  \tikzstyle{node}=[circle,fill=none,draw=black, minimum size=4mm]
  \node[state]         (s)               {$s$};
  \node[node]         (i1)[below left of=s, yshift=-5mm]   {};
  \node[node]         (i2)[below right of=s,yshift=-5mm]   {};
  \node[node]         (i3)[right of=i2]   {};
  \node[node]         (i4)[left  of=i1]   {};
  \node[node]         (j1)[below of=i1]   {};
  \node[node]         (j2)[below of=i2]   {};
  \node[node]         (j3)[below of=i3]   {};
  \node[node]         (j4)[below of=i4]   {};
  \node[state]         (t)[below right of=j1,yshift=-5mm]   {$t$};
\path (s)   edge [bend left]       node {} (i2)
            edge [bend left]       node {} (i3)
            edge [bend right]      node {} (i1)
            edge [bend right]      node {} (i4)
       (i1) edge                   node {} (j1)
       (i2) edge                   node {} (j2)
       (i3) edge                   node {} (j3)
       (i4) edge                   node {} (j4)
       (j2) edge [bend left]       node {} (t)
       (j3) edge [bend left]       node {} (t)
       (j1) edge [bend right]      node {} (t)
       (j4) edge [bend right]      node {} (t);
\end{tikzpicture}
\caption{Network $G_3$ with $|T|=1$, $|J|=8$, $C_t=4$.}
\label{fig_eg2}
\end{center}
\end{figure}
\begin{itemize}
  \item In the case $d=0$, obviously,
  \begin{align*}
  &2+\left[ \sum_{t\in T}|R_t(\dt-d)|{d+|J|+1 \choose |J|} \right]^{\frac{1}{d+1}}\\
  =&2+9|R_t(2)|>|R_t(2)|.
  \end{align*}
  \item In the case $d=1$, after a simple calculation, we deduce that
  $$\sum_{t\in T}|R_t(\dt-d)|=|R_t(1)|=12,$$ and
  \begin{equation*}
  \begin{split}
  &2+\left[ \sum_{t\in T}|R_t(\dt-d)|{d+|J|+1 \choose |J|} \right]^{\frac{1}{d+1}}\\
  =&2+\left[ |R_t(1)|{1+8+1\choose 8}\right]^{\frac{1}{2}}=2+(12\times45)^{\frac{1}{2}}\geq 20.
  \end{split}
  \end{equation*}
  Therefore,
  \begin{align*}
  &2+\left[ \sum_{t\in T}|R_t(\dt-d)|{d+|J|+1 \choose |J|} \right]^{\frac{1}{d+1}}\\
  >&\sum_{t\in T}|R_t(\dt-d)|.
  \end{align*}
\end{itemize}
\end{eg}
\section{Conclusions}
In this paper, using the extended global encoding kernels proposed by Zhang in \cite{zhang-correction}, we can prove the refined Singleton bound in network error correction coding more easily, and give a constructive proof to show that this bound is tight, that is, we construct network MDS codes which meet this bound with equality. As a consequence of this proof, an algorithm is designed to construct linear network error correction codes, especially network MDS codes. The time complexity of the proposed algorithm is analyzed. It is shown that the required field size for the existence of linear network error correction codes can become smaller than the previously known results, and even much smaller in some cases.

For random linear network error correction coding, the upper bounds on the failure probabilities for network MDS codes and general linear network error correction codes are obtained. And we slightly improve on the probability mass function of the minimum distance of the random linear network error correction codes introduced in \cite{zhang-random}, as well as the upper bound on the field size required for the existence of linear network error correction codes with degradation at most $d$.

\appendices
\section{Proof of Theorem \ref{thm_compare_size}}\label{app}
\begin{IEEEproof}
We choose an error pattern $\p_1\in R_t(\bt)$ arbitrarily, that is, the chosen error pattern $\p_1$ satisfies $|\p_1|=rank_t(\p_1)=\bt$. Then we can extend $\p_1$ to an error pattern $\p_1'$ with $\p_1\subseteq \p_1'$ and $|\p_1'|=rank_t(\p_1')=C_t$, since the minimum cut capacity between $s$ and $t$ is $C_t$. Define two sets as follows:
$$\Omega_{1,\bt}=\{\mbox{error pattern } \p\subseteq \p_1': \p\in R_t(\bt) \}$$
and
$$\Omega_{1,\dt}=\{\mbox{error pattern } \p'\subseteq \p_1': \p'\in R_t(\dt) \}.$$
From the above definitions, we have
$$|\Omega_{1,\bt}|={C_t \choose \bt}\ \mbox{and}\ |\Omega_{1,\dt}|={C_t \choose \dt}.$$
Note that $\bt\leq \dt \leq \lfloor\frac{C_t}{2}\rfloor$ implies ${C_t \choose \bt}\leq {C_t \choose \dt}$. In other words, for each $\p\in \Omega_{1,\bt}$, there exists an error pattern $\p' \in \Omega_{1,\dt}$ such that $\p$ is covered by $\p'$, i.e., $\p\subseteq\p'$, and $\theta'\neq \eta'$ for any distinct $\theta,\eta\in \Omega_{1,\bt}$.

Again, choose an error pattern $\p_2 \in R_t(\bt)\backslash \Omega_{1,\bt}$ arbitrarily. In the same way as for $\p_1$, $\p_2$ can be extended to an error pattern $\p_2'$ with $\p_2\subseteq \p_2'$ and $|\p_2'|=rank_t(\p_2')=C_t$. Define the next two sets:
$$\Omega_{2,\bt}=\{\mbox{error pattern } \p\subseteq \p_2': \p\in R_t(\bt), \p\nsubseteq \p_1'\cap \p_2' \},$$
and
$$\Omega_{2,\dt}=\{\mbox{error pattern } \p'\subseteq \p_2': \p'\in R_t(\dt), \p'\nsubseteq \p_1'\cap \p_2' \}.$$
Obviously, for all $\p\in \Omega_{2,\bt}$ and $\p'\in \Omega_{2,\dt}$, we have $\p\notin \Omega_{1,\bt}$ and $\p'\notin \Omega_{1,\dt}$. This means that $\Omega_{1,\bt}\cap\Omega_{2,\bt}=\emptyset$ and
$\Omega_{1,\dt}\cap\Omega_{2,\dt}=\emptyset$. Let $|\p_1'\cap\p_2'|=k_{1,2}$. Then
$$|\Omega_{2,\bt}|={C_t\choose \bt}-{k_{1,2}\choose \bt}\mbox{ and }|\Omega_{2,\dt}|={C_t\choose \dt}-{k_{1,2}\choose \dt}.$$
We adopt the convention that ${a\choose b}=0$ for $a<b$.

Similarly, we choose an error pattern $\p_3\in R_t(\bt)\backslash \Omega_{1,\bt}\cup \Omega_{2,\bt}$, and extend $\p_3$ to an error pattern $\p_3'$ with $\p_3\subseteq \p_3'$ and $|\p_3'|=rank_t(\p_3')=C_t$. Define
$$\Omega_{3,\bt}=\{\p\subseteq \p_3': \p\in R_t(\bt), \p\nsubseteq \{\p_1'\cup \p_2'\}\cap\p_3' \},$$
and
$$\Omega_{3,\dt}=\{\p'\subseteq \p_3': \p'\in R_t(\dt), \p'\nsubseteq \{\p_1'\cup \p_2'\}\cap\p_3' \}.$$
We claim that for all $\p\in \Omega_{3,\bt}$ and $\p'\in \Omega_{3,\dt}$, $\p\notin \Omega_{1,\bt}\cup \Omega_{2,\bt}$ and $\p'\notin \Omega_{1,\dt}\cup\Omega_{2,\dt}$. Conversely, suppose that $\p \in \cup_{i=1}^2 \Omega_{i,\bt}$ (resp. $\p' \in \cup_{i=1}^2 \Omega_{i,\dt}$). Together with $\p\in \Omega_{3,\bt}$ (resp. $\p'\in \Omega_{3,\dt}$), this shows that $\p\subseteq \{\p_1'\cup\p_2'\}\cap\p_3'$ (resp. $\p'\subseteq \{\p_1'\cup\p_2'\}\cap\p_3'$). It contradicts to our choice $\p\in \Omega_{3,\bt}$. Thus, $\Omega_{3,\bt}\cap\Omega_{i,\bt}=\emptyset$ and $\Omega_{3,\dt}\cap\Omega_{i,\dt}=\emptyset,\ i=1,2.$
Further, let $|\{\p_1'\cup\p_2'\}\cap\p_3'|=k_{1,2,3}$. Then
$$|\Omega_{3,\bt}|={C_t\choose \bt}-{k_{1,2,3}\choose \bt}\mbox{ and }|\Omega_{3,\dt}|={C_t\choose \dt}-{k_{1,2,3}\choose \dt}.$$

Choose an error pattern $\p_4\in R_t(\bt)\backslash \cup_{i=1}^3\Omega_{i,\bt}$, and extend $\p_4$ to an error pattern $\p_4'$ with $\p_4\subseteq \p_4'$ and $|\p_4'|=rank_t(\p_4')=C_t$. Define two sets similarly:
$$\Omega_{4,\bt}=\{\p\subseteq \p_4': \p\in R_t(\bt), \p\nsubseteq \{\cup_{i=1}^3\p_i'\}\cap\p_4' \},$$
and
$$\Omega_{4,\dt}=\{\p'\subseteq \p_4': \p'\in R_t(\dt), \p'\nsubseteq \{\cup_{i=1}^3\p_i'\}\cap\p_4' \}.$$
For all $\p\in \Omega_{4,\bt}$ and $\p'\in \Omega_{4,\dt}$, $\p\notin \cup_{i=1}^3\Omega_{i,\bt}$ and $\p'\notin \cup_{i=1}^3\Omega_{i,\dt}$. Assume the contrary, i.e., $\p \in \cup_{i=1}^3 \Omega_{i,\bt}$, which implies that $\p\subseteq \{\p_1'\cup\p_2'\cup\p_3'\}\cap\p_4'$. It is a contradiction. Similarly, we have $\p'\notin \cup_{i=1}^3\Omega_{i,\dt}$ for all $\p'\in \Omega_{4,\dt}$.
That is, $\Omega_{4,\bt}\cap\Omega_{i,\bt}=\emptyset$ and $\Omega_{4,\dt}\cap\Omega_{i,\dt}=\emptyset,\ i=1,2,3.$ Let $|\{\cup_{i=1}^3\p_i'\}\cap\p_4'|=k_{1,2,3,4}$. It follows that
$$|\Omega_{4,\bt}|={C_t\choose \bt}-{k_{1,2,3,4}\choose \bt},\ |\Omega_{4,\dt}|={C_t\choose \dt}-{k_{1,2,3,4}\choose \dt}.$$

We continue this procedure until we cannot choose a new error pattern $\p\in R_t(\bt)$. Since $|R_t(\bt)|$ is finite, this procedure will stop at some step. Without loss of generality, assume that the procedure stops at the $m$th step. That is, $R_t(\bt)=\cup_{i=1}^m\Omega_{i,\bt}$. Together with what we have proved above, $\Omega_{i,\bt}\cap\Omega_{j,\bt}=\emptyset$ for all $i,j$ satisfying $i\neq j$ ($1\leq i,j \leq m)$. This implies that
$$|R_t(\bt)|=\sum\limits_{i=1}^m|\Omega_{i,\bt}|=\sum_{i=1}^m\left[{C_t\choose \bt}-{k_{1,2,\cdots,i}\choose \bt}\right],$$
where set $k_1=0$. Similarly, we also have
$\Omega_{i,\dt}\cap\Omega_{j,\dt}=\emptyset$ for all $i,j$ satisfying $i\neq j$ ($1\leq i,j \leq m$), and $\cup_{i=1}^m\Omega_{i,\dt}\subseteq R_t(\dt)$, which implies that
$$|R_t(\dt)|\geq\sum\limits_{i=1}^m|\Omega_{i,\dt}|=\sum\limits_{i=1}^m\left[{C_t\choose \dt}-{k_{1,2,\cdots,i}\choose \dt}\right].$$

In order to prove $|R_t(\bt)|\leq|R_t(\dt)|$, it suffices to show $|\Omega_{i,\bt}|\leq|\Omega_{i,\dt}|$, i.e., ${C_t\choose \bt}-{k_{1,2,\cdots,i}\choose \bt}\leq{C_t\choose \dt}-{k_{1,2,\cdots,i}\choose \dt}$ for each $i=1,2,\cdots,m$.

To simplify the notation, we omit the subscripts in the following discussion. It follows that we just need to prove
$${C\choose \delta}-{k \choose \delta}\geq{C\choose \beta}-{k \choose \beta},$$
that is,
\begin{eqnarray}\label{choose}
{C\choose \delta}-{C\choose \beta}\geq{k \choose \delta}-{k \choose \beta},
\end{eqnarray}
where $\beta\leq \delta \leq \lfloor \frac{C}{2} \rfloor$ and $k\leq C$.

If $k<\delta$, the inequality (\ref{choose}) immediately holds.
Otherwise $k\geq \delta$, note that
\begin{align*}
&{C\choose \delta}-{C \choose \beta}\\
=&\left[{C\choose \delta}-{C \choose \delta-1}\right]+\left[{C\choose \delta-1}-{C \choose \delta-2}\right]+\cdots\\
&+\left[{C\choose \beta+2}-{C \choose \beta+1}\right]+\left[{C\choose \beta+1}-{C \choose \beta}\right],
\end{align*}
and
\begin{align*}
&{k\choose \delta}-{k \choose \beta}\\
=&\left[{k \choose \delta}-{k \choose \delta-1}\right]+\left[{k \choose \delta-1}-{k \choose \delta-2}\right]+\cdots\\
&+\left[{k \choose \beta+2}-{k \choose \beta+1}\right]+\left[{k \choose \beta+1}-{k \choose \beta}\right].
\end{align*}
This implies that the inequality (\ref{choose}) holds provided that we can show
$${C\choose a+1}-{C \choose a}\geq{k\choose a+1}-{k \choose a}$$
for any $a$ satisfying $\beta \leq a \leq \delta-1$.
After a simple calculation, it is equivalent to prove
\begin{align}\label{choose2}
&C(C-1)\cdots(C-a+1)(C-2a-1)\nonumber\\
\geq& k(k-1)\cdots(k-a+1)(k-2a-1).
\end{align}
It is not difficult to see that the inequality (\ref{choose2}) holds for $k \geq \delta$.
This completes the proof.
\end{IEEEproof}


\end{document}